\documentclass[amsmath,amssymb,aps,prd,nofootinbib,twocolumn,superscriptaddress]{revtex4-2}
\usepackage[utf8]{inputenc}
\usepackage{mathrsfs}
\usepackage{bm}
\usepackage{url}
\usepackage[normalem]{ulem}
\usepackage{mathtools}
\usepackage{array}
\newcolumntype{P}[1]{>{\centering\arraybackslash}p{#1}}
\newcolumntype{M}[1]{>{\centering\arraybackslash}m{#1}}
\usepackage[caption=false]{subfig}
\usepackage{dcolumn}
\usepackage{graphicx, epsfig}
\usepackage[dvipsnames]{xcolor}
\usepackage{mathrsfs}
\usepackage{bm}
\usepackage{yhmath}
\usepackage[caption=false]{subfig}
\usepackage[normalem]{ulem}
\usepackage{mathtools}
\usepackage{bigints}
\usepackage{float}
\usepackage[colorlinks = true, linkcolor = purple, urlcolor  = blue, citecolor = blue, anchorcolor = blue]{hyperref}
\usepackage{float}
\usepackage{multirow}

\def\apj{{ Astrophys. J.}}
\def\apjl{{ Astrophys. J. Lett.}}

\def\mnras{{ Mon. Not. Roy. Astron. Soc.}}

\def\nat {{ Nature}}

\begin{document}

\title{Two-component off-axis jet model for radio flares of tidal disruption events}

\author{Yuri Sato}
\email{yuris@phys.aoyama.ac.jp (YS)}
\affiliation{Department of Physical Sciences, Aoyama Gakuin University, 5-10-1 Fuchinobe, Sagamihara 252-5258, Japan}
\author{Kohta Murase}
\email{murase@psu.edu (KM)}
\affiliation{Department of Physics; Department of Astronomy \& Astrophysics; Center for Multimessenger Astrophysics, Institute for Gravitation and the Cosmos, The Pennsylvania State University, University Park, PA 16802, USA}
\affiliation{Center for Gravitational Physics and Quantum Information, Yukawa Institute for Theoretical Physics, Kyoto University, Kyoto, Kyoto 606-8502, Japan}
\affiliation{School of Natural Sciences, Institute for Advanced Study, Princeton, New Jersey 08540, USA}
\author{Mukul Bhattacharya}
\affiliation{Department of Physics; Department of Astronomy \& Astrophysics; Center for Multimessenger Astrophysics, Institute for Gravitation and the Cosmos, The Pennsylvania State University, University Park, PA 16802, USA}
\affiliation{Department of Physics, Wisconsin IceCube Particle Astrophysics Center, University of Wisconsin, Madison, WI 53703, USA}
\author{Jose A. Carpio}
\affiliation{Department of Physics \& Astronomy; Nevada Center for Astrophysics, University of Nevada, Las Vegas, NV 89154, USA}
\author{Mainak Mukhopadhyay}
\affiliation{Department of Physics; Department of Astronomy \& Astrophysics; Center for Multimessenger Astrophysics, Institute for Gravitation and the Cosmos, The Pennsylvania State University, University Park, PA 16802, USA}
\author{B. Theodore Zhang}
\affiliation{Center for Gravitational Physics and Quantum Information, Yukawa Institute for Theoretical Physics, Kyoto University, Kyoto, Kyoto 606-8502, Japan}            

\begin{abstract}
Recently, radio emission from tidal disruption events (TDEs) has been observed from months to years after the optical discovery. Some of the TDEs including ASASSN-14ae, ASASSN-15oi, AT~2018hyz, and AT~2019dsg are accompanied by the late-time rebrightening phase characterized by a rapid increase in the radio flux. 
We show that it can be explained by the off-axis two-component jet model, in which the late-time rebrightening arises from the off-axis view of a decelerating narrower jet with an initial Lorentz factor of $\sim10$ and a jet opening angle of $\sim0.1\,{\rm rad}$, while the early-time radio emission is attributed to the off-axis view of a wider jet component. We also argue that the rate density of jetted TDEs inferred from these events is consistent with the observations.
\end{abstract}


\maketitle

\section{Introduction} 
A tidal disruption event (TDE) occurs when a star is torn apart by the tidal forces of a supermassive black hole (SMBH) at sufficiently close approaches \citep{Hills1975,Rees1988,EK1989}.
TDEs provide rich multiwavelength data in radio, infrared (IR)/optical/ultraviolet (UV), x- and gamma-ray bands, 
which can be used to study the properties of quiescent galaxies, jets, and the circumnuclear medium (CNM). 
While thermal emission is believed to arise from the debris of the disrupted star 
by reprocessing \citep{Loeb1997,Strubbe2009,Lodato2011,Komossa2015,MS2016,Roth2016}, 
details of the mechanism remain unclear.
Nonthermal emission has also been observed for some TDEs exhibiting outflows, and radio emission is attributed to synchrotron emission from relativistic electrons accelerated in jets or winds. Future multifrequency radio observations will help us reveal the structure and evolution of the outflows \citep{DeColle2012,Alexander2020,van_Velzen2021}.

At least some TDEs can launch relativistic jets, as inferred from variable x-ray and subsequent radio emission, and 4 jetted TDEs (Swift J1644+57 \citep{Bloom2011,Burrows2011}, Swift J2058+05 \citep{Cenko2012}, Swift J1112-8238 \citep{Brown2015} and AT~2022cmc \citep{Andreoni2022}) are known to date. 
The apparent rate density of jetted TDEs is $\sim 0.03~{\rm Gpc^{-3}yr^{-1}}$ \citep{Burrows2011,Andreoni2022}, which is only $\lesssim 1\%$ of the total TDE rate density, $\sim 10^{2-3}\,{\rm Gpc}^{-3}\,{\rm yr}^{-1}$ \citep{Teboul2022}. 
In addition to these jetted events, a growing sample of optically-detected TDEs exhibit radio afterglows that are consistent with emission from nonrelativistic outflows (see Ref.~\citep{Alexander2020} for a review). 

The recent discovery of late-time radio emission from some TDEs (e.g., ASASSN-14ae \citep{Holoien2014}, AT~2018hyz \citep{Cendes2022}, AT~2019azh \citep{Goodwin2022}) and radio rebrightening in some other TDEs (e.g., iPTF16fnl \citep{Horesh2021b}, AT~2019dsg \citep{Stein2021}, ASASSN-15oi \citep{Horesh2021}) point to outflows that become observable after a significant delay (from a few months to years) since the disruption. 
In particular, Ref.~\citep{Cendes2023} reported late-time radio brightening in some TDEs on a timescale of $\sim 2-4\,{\rm years}$, where more than half of the sample still shows rising emission in the radio band. These TDEs showed a rapid rise and rebrightening in radio emission at late times with a peak radio luminosity of $\sim10^{38}-10^{39}~{\rm erg~s^{-1}}$. The underlying mechanism of these delayed radio flares in TDEs is of interest, and the outflows can be either jets \citep{Giannios2011,Piran2015,van_Velzen2016,Matsumoto:2022hqp,Beniamini2023,Sfaradi2023} or winds \citep{Alexander2016,Matsumoto:2021qqo,Cendes2021,Hayasaki2023}.

A relativistic jet launched from a black hole -- accretion disk system is expected to be structured, as demonstrated in studies of gamma-ray bursts (GRBs) both theoretically and observationally (see e.g., Refs.~\citep{Rossi2002,Zhang2002,Alexander2018,Ioka2019,Gottlieb2020,Zhang:2023uei,Obayashi2024}). 
Such a structured jet is often ``modeled'' as a two-component jet with a faster inner component and slower outer component, and has been exploited to explain multiwavelength data of afterglows (see e.g., Refs.~\citep{Berger2003,Racusin2008,Sato2021,Sato2023a,Sato2023b} for GRBs and Refs.~\cite{Wang:2014nwa,Mimica2015,Liu2015,Zhou2023,Teboul2023,Yuan2024} for TDEs).
In this work, we explore the origin of late radio flares such as the rapid rising part and rebrightening of ASASSN-14ae, ASASSN-15oi, AT~2018hyz, and AT~2019dsg, with luminosities of $\sim10^{38}-10^{39}~{\rm erg~s^{-1}}$, and propose a two-component off-axis jet model for these 4 TDEs.
We show that this model provides a natural explanation for both the observed radio rebrightening at late times and the earlier radio data without the need for late-time engine activity.

\begin{table*}
\caption{Parameters used for modeling 4 radio TDEs: ASASSN-14ae, ASASSN-15oi, AT~2018hyz, and AT~2019dsg. 
}
\label{tab:parameter}
\scalebox{0.85}{
\begin{tabular}{llcccccccccc}
\hline
& & ~~$\theta_v$~[rad]~~ & ~~$\theta_j$~[rad]~~ & ~~${\mathcal E}_{k}^{\rm iso}$~[erg]~~ & ~~$\Gamma_0$~~ & ~~$n_{\rm ext}$~${[\rm cm^{-3}]}$~~ & ~~$w$~~ & ~~$s$~~ & ~~$\epsilon_B$~~ & ~~$\epsilon_e$~~ & ~~$f_e$~~  \\
\hline
\multirow{4}{*}{\textbf{Narrow jet}} & ASASSN-14ae & \multirow{4}{*}{$1.2$} & \multirow{4}{*}{$0.11$} & \multirow{4}{*}{$8.0\times10^{54}$} & \multirow{4}{*}{$10$} & $1.0$ & $1.0$ & $2.6$ & $3.0\times 10^{-4}$ & $0.1$ & $0.1$  \\
 & ASASSN-15oi & & & & & $2.0$ & $0.5$ & $2.8$ & $3.0\times 10^{-3}$ & $0.06$ & $0.8$  \\
& AT~2018hyz & & & & & $10.0$ & $0.5$ & $2.8$ & $2.0\times 10^{-4}$ & $0.1$ & $0.1$  \\
& AT~2019dsg & & & & & $3.0$ & $1.0$ & $2.5$ & $1.0\times 10^{-3}$ & $0.1$ & $0.1$  \\
\hline 
\multirow{4}{*}{\textbf{Wide jet}} & ASASSN-14ae &\multirow{4}{*}{$1.2$} & \multirow{4}{*}{$0.34$} & \multirow{4}{*}{$1.0\times10^{52}$} & \multirow{4}{*}{$3$}& $1.0$ & $1.0$ & $2.5$ & $2.0\times 10^{-3}$ & $0.1$ & $0.1$  \\
 & ASASSN-15oi & & & & & $2.0$ & $0.5$ & $2.8$ & $5.0\times 10^{-3}$ & $0.2$ & $0.3$ \\
& AT~2018hyz & & & & & $10.0$ & $0.5$ & $2.8$ & $1.5\times 10^{-4}$ & $0.1$ & $0.1$  \\
& AT~2019dsg & & & & & $3.0$ & $1.0$ & $2.8$ & $1.0\times 10^{-2}$ & $0.1$ & $0.1$  \\
\hline
\end{tabular}}
\end{table*}

\section{One-component jet model}
\label{sec:one-jet}

We first consider the standard one-component jet model viewed off-axis to discuss the rapid rise in the radio flux at late times, considering 4 TDEs, namely ASASSN-14ae, ASASSN-15oi, AT~2018hyz, and AT~2019dsg.  
The jet has an half-opening angle $\theta_j$, initial Lorentz factor $\Gamma_0$, isotropic kinetic energy ${\mathcal E}_k^{\rm iso}$, and viewing angle $\theta_v$ measured from the jet axis. As the jet propagates through an external medium with a power-law density profile, $n(r) =n_{\rm ext} (r/r_{\rm ext})^{-w}$ with $r_{\rm ext}=10^{18}~{\rm cm}$, external shocks are formed. 
Focusing on the forward shock, we numerically solve the blast wave radius $r(t)$ and its Lorentz factor $\Gamma(t)$, for an initial radius of $10^{13}\,{\rm cm}$. 
This radius is comparable to the tidal disruption radius,  
and our results are unaffected by the choice of $r(t=0)$. 

To obtain the radio light curves, we utilize the afterglow module of {\sc AMES} (Astrophysical Multimessenger Emission Simulator), following the treatments described in Refs.~\cite{Zhang2021,Zhang:2023uei} (see also Ref.~\cite{Murase:2010fq}). Electrons are assumed to be accelerated via the diffusive shock acceleration mechanism, resulting in an electron injection spectrum $\varepsilon_e^{-s}$, where $s$ is the spectral index. A fraction $\epsilon_B$ of the downstream internal energy density of the shocked material is converted to the magnetic field, while a fraction $\epsilon_e$ is carried by nonthermal electrons. The electron spectrum is calculated by solving the kinetic equation in the no escape limit, as outlined in Ref.~\cite{Zhang:2023uei}, accounting for synchrotron, inverse-Compton, and adiabatic loss processes. 
For a luminosity distance $d_L$, considering the equal-arrival-time-surface (EATS), the observed photon flux at 
time $T=t-t_0$ since the time of discovery ($t_0$) and at frequency $\nu$ is calculated as \citep{Takahashi2022,Zhang:2023uei}
\small
\begin{align}\label{flux-EATS} 
    F_\nu(T) &= \frac{1+z}{d_L^2} \int_0^{\theta_j} d\theta \, {\rm sin}\theta \int_0^{2\pi} d\phi \frac{r^2 |\mu - \beta_{\rm sh}|}{1-\mu \beta_{\rm sh}} \nonumber \\ &\times \frac{1}{\Gamma^2{(1-\beta\mu)}^2}\frac{j'_{\nu'}}{\alpha_{\nu}} (1 - e^{-\tau_\nu})|_{\hat{t} = (T+t_0)/(1+z)+ \mu r / c},
\end{align}
\normalsize
where $\beta_{\rm sh} = \sqrt{1-\Gamma_{\rm sh}^{-2}}$ is the shock velocity and its Lorentz factor is $\Gamma_{\rm sh}\approx\sqrt{2}\Gamma$. The integration variables $\theta$ and $\phi$ are the polar angle ($\theta=0$ is the jet axis) and azimuthal angle, respectively, and $\mu = {\rm sin}\theta {\rm sin} \theta_v {\rm cos} \phi + {\rm cos} \theta {\rm cos} \theta_v$. Here, $j'_{\nu'}$ is the comoving emission coefficient, and $\alpha_{\nu}$ is the absorption coefficient in the engine frame (that is dropped when the attenuation is irrelevant).
For late-time radio observations, $t_0$ can be assumed to be negligibly small, which implies that the disk is formed and the jet is launched instantaneously after the tidal disruption occurs at $t=0$. 
As we consider delayed radio emission at a significantly later epoch of $T \sim 10^7-10^8\,{\rm s}$, the light curves obtained here are not affected.

\begin{figure*}
\begin{minipage}{0.4\linewidth}
\centering
\includegraphics[width=1.0\textwidth]{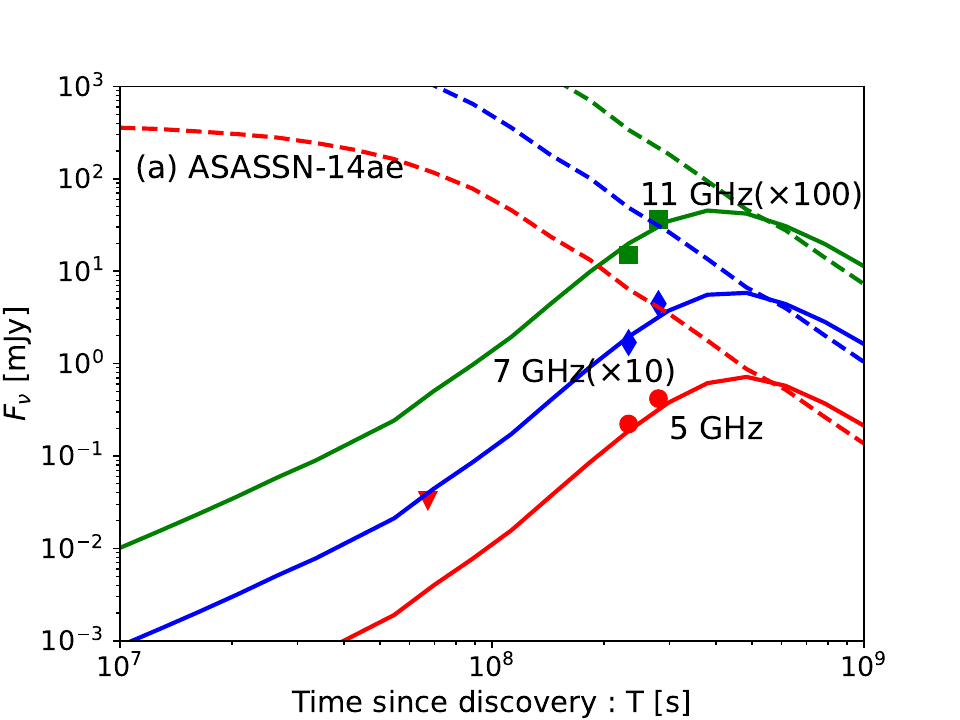}
\end{minipage}
\begin{minipage}{0.4\linewidth}
\centering
\includegraphics[width=1.0\textwidth]{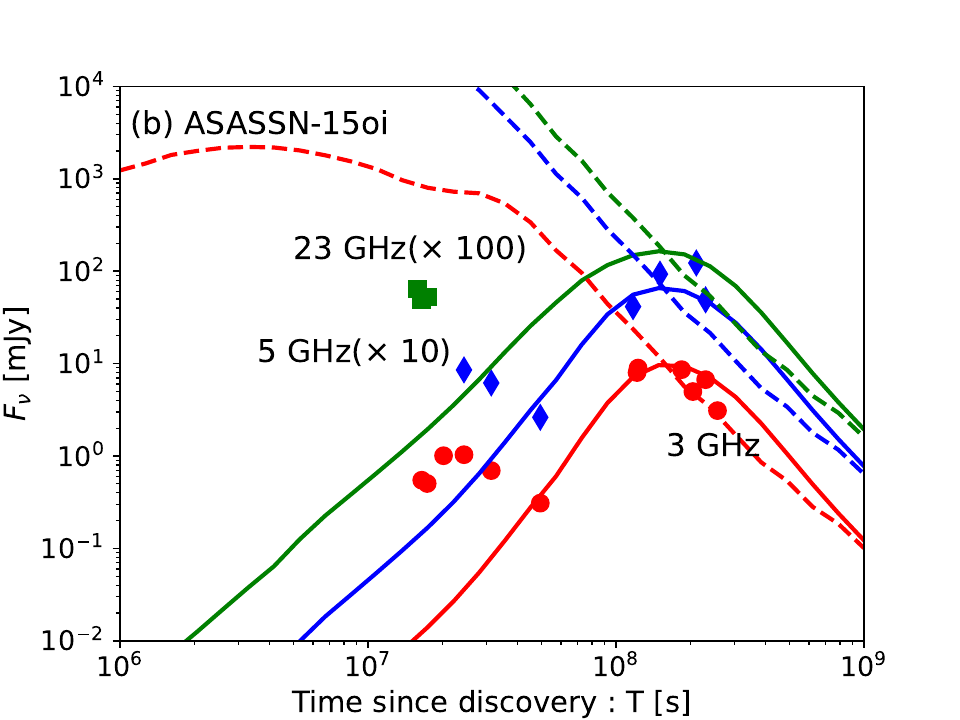}
\end{minipage}
\begin{minipage}{0.4\linewidth}
\centering
\includegraphics[width=1.0\textwidth]{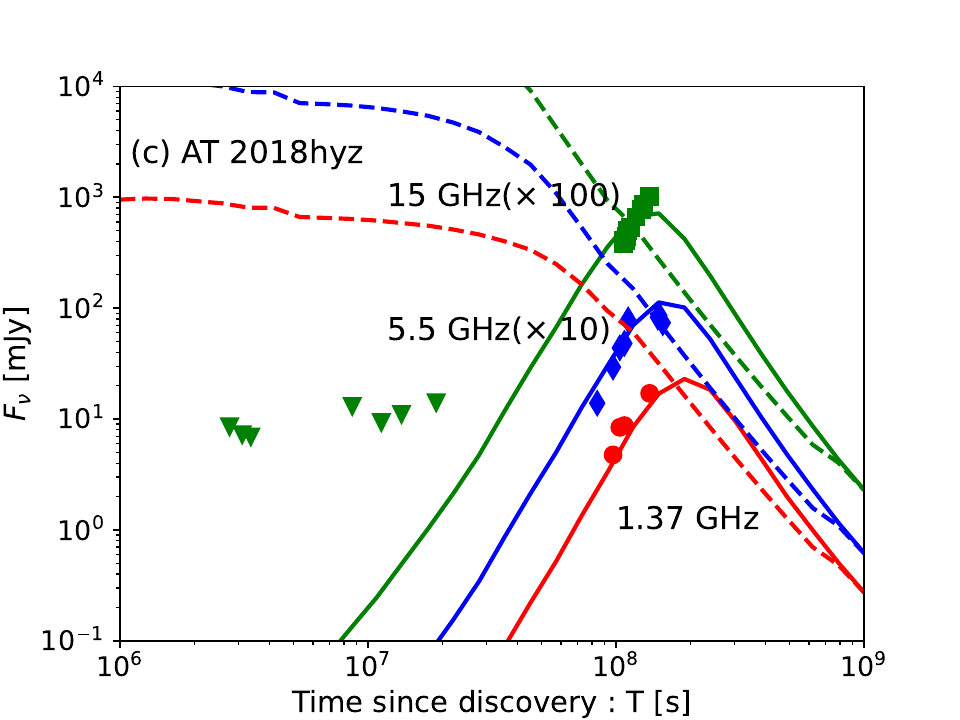}
\end{minipage}
\begin{minipage}{0.4\linewidth}
\centering
\includegraphics[width=1.0\textwidth]{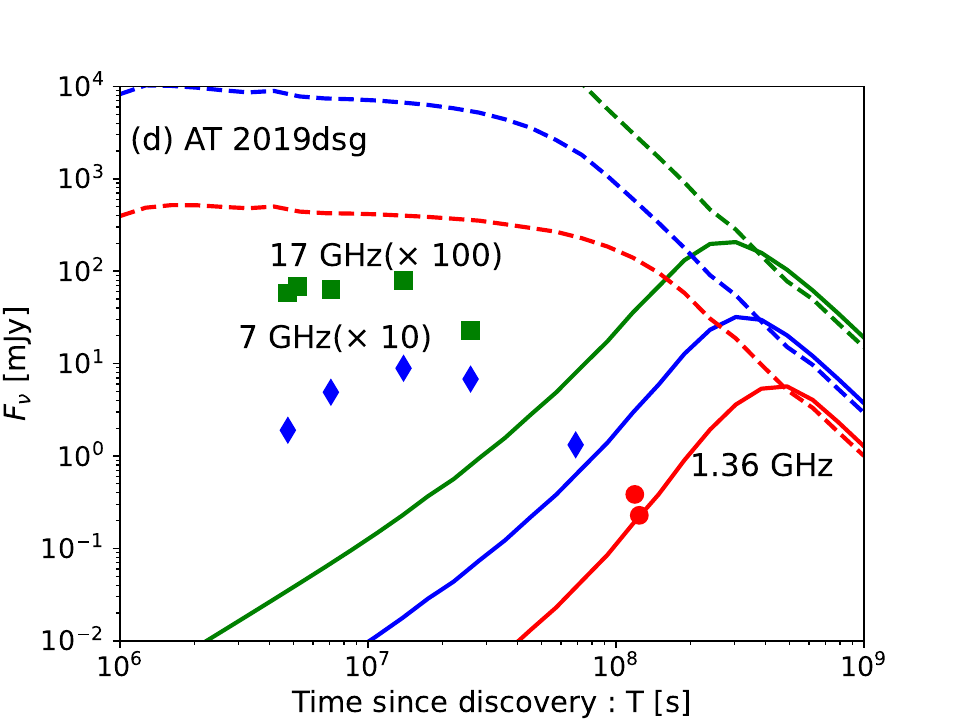}
\end{minipage}
\vspace{-0.1cm}
\caption{
Observed radio data for ASASSN-14ae (5~GHz: red filled-circles, 7~GHz: blue diamonds, and 11~GHz: green squares) taken from Ref.~\citep{Cendes2023}, ASASSN-15oi (3~GHz: red filled-circles, 5~GHz: blue diamonds, and 23~GHz: green squares) obtained from  Refs.~\citep{Horesh2021,Anumarlapudi2024,Hajela2024}, AT~2018hyz (1.37~GHz: red filled-circles, 5.5~GHz: blue diamonds, and 15~GHz: green squares) extracted from Refs.~\citep{Cendes2022,Anumarlapudi2024} and AT~2019dsg (1.36~GHz: red filled-circles, 7~GHz: blue diamonds, and 17~GHz: green squares) derived from Refs.~\citep{Cendes2021,Cendes2023} are shown. These are compared
with the light curves from single jets in the radio bands
[ASASSN-14ae; 5~GHz: red, 7~GHz: blue, and 11~GHz: green), ASASSN-15oi (3~GHz: red, 5~GHz: blue, and 23~GHz: green), AT~2018hyz (1.37~GHz: red, 5.5~GHz: blue, and 15~GHz: green) and AT~2019dsg (1.36~GHz: red, 7~GHz: blue, and 17~GHz: green].
The upper limits in radio flux are shown with downward triangles.
In all four panels, the solid lines represent the emission from our narrow jet viewed off-axis, and the dashed lines show the results for the on-axis viewing case ($\theta_v=0$ with other parameters unchanged).
}
\label{one-jet}
\end{figure*}

Radio light curves from our theoretical model are shown with solid lines in Figs~\ref{one-jet} (a)-(d), and the model parameters are presented in Table~\ref{tab:parameter}.
Our narrow jet explains a rapid rise in radio bands at the late epoch for all 4 TDEs through relativistic beaming effect if the jet is viewed off-axis \citep{Granot2002}.
The beaming effect becomes weaker as the jet decelerates, resulting in the rising behavior, and the flux peaks when $\Gamma\sim(\theta_v-\theta_j)^{-1}$, after which the emission asymptotically approaches the on-axis light curve ($\theta_v=0$: dashed lines in Figs.~\ref{one-jet} (a)-(d)) \citep{Granot2002}.
At $T\sim10^6-10^9\,{\rm s}$, we find that the absorption frequency $\nu_a$, the typical frequency $\nu_m$, and the cooling frequency $\nu_c$ are ordered as $\nu_a<\nu_m<\nu_c$, where $\nu_a\sim10^9\,{\rm Hz}$ and $\nu_m\sim10^{10}-10^{12}\,{\rm Hz}$ in our cases. 
The value of $\nu_m$ ($\nu_a$) is larger (smaller) than the radio bands. In the off-axis case, the light curves evolve as $F_{\nu} \propto T^{(21-8w)/3}$.
We adopt $w=0.5$ for ASASSN-15oi and AT~2018hyz, and $w=1.0$ for ASASSN-14ae and AT~2019dsg.
The radio light curves at late times follow $F_{\nu} \approx T^{5.7}$ and $\approx T^{4.3}$ for $w=0.5$ and $1.0$, respectively, which are consistent with the observed data.
For comparison, the dashed lines in Figs.~\ref{one-jet} (a)-(d) show the light curves in the on-axis viewing case,
for which the flux decreases at later times. 
In the off-axis case, the flux starts with a rising part due to the relativistic beaming effect \citep{Granot2002}. 
The radio data of ASASSN-14ae and AT~2018hyz can be explained with such an off-axis jet (see also Refs.~\citep{Matsumoto2023,Beniamini2023,Sfaradi2023}).
However, for ASASSN-15oi and AT~2019dsg, the observations for $T\lesssim 10^8\,{\rm s}$ shows another declining phase before the radio rebrightening phase. 
The radio emission from our narrow jet viewed off-axis is inconsistent with the earlier radio data, making it difficult for the one-component jet model to describe all the radio data. 

\section{Two-Component Jet Model}
\label{sec:two-jet}

\begin{figure*}
\begin{minipage}{0.4\linewidth}
\centering
\includegraphics[width=1.0\textwidth]{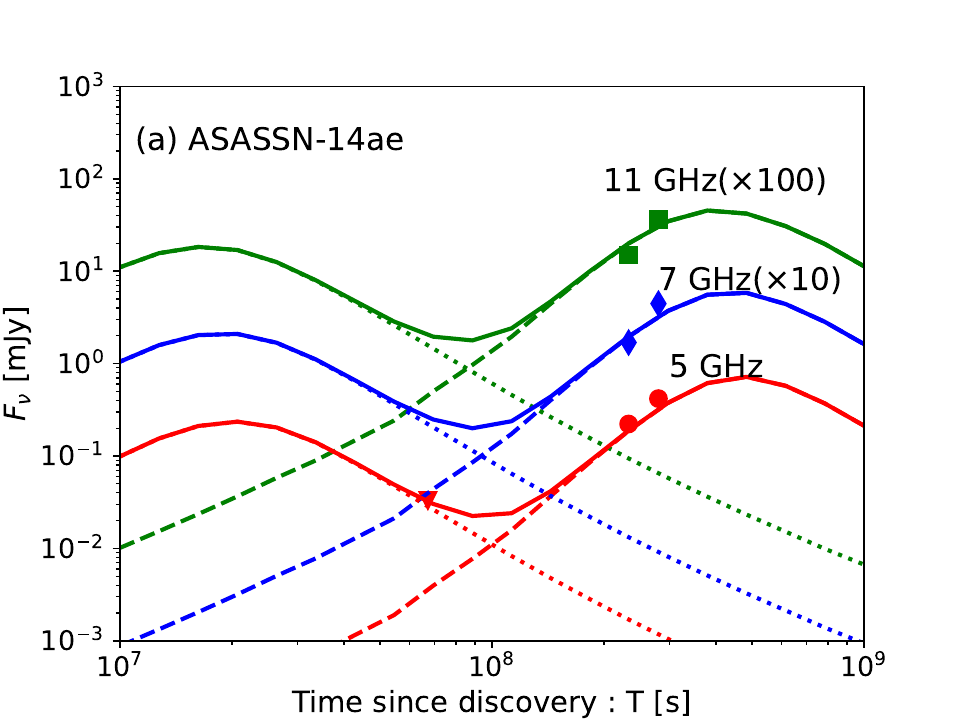}
\end{minipage}
\begin{minipage}{0.4\linewidth}
\centering
\includegraphics[width=1.0\textwidth]{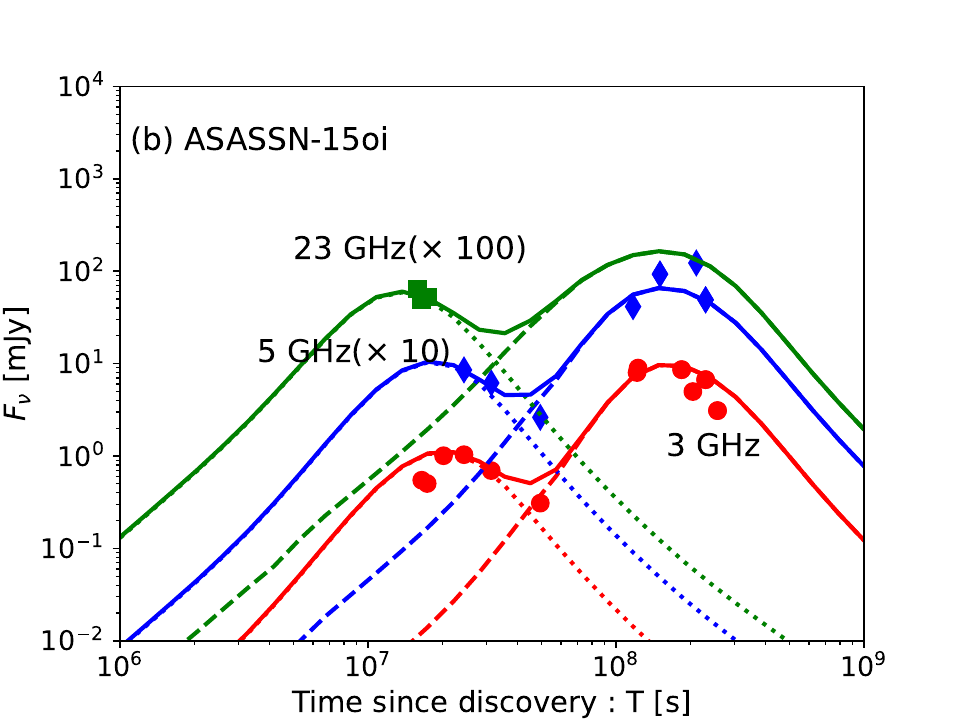}
\end{minipage}
\begin{minipage}{0.4\linewidth}
\centering
\includegraphics[width=1.0\textwidth]{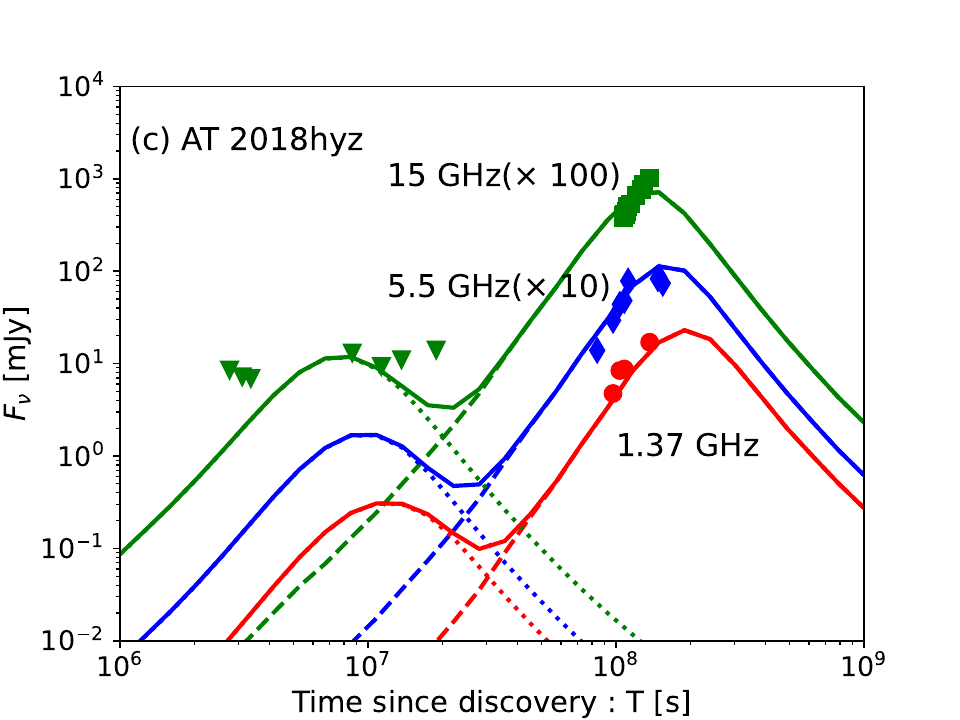}
\end{minipage}
\begin{minipage}{0.4\linewidth}
\centering
\includegraphics[width=1.0\textwidth]{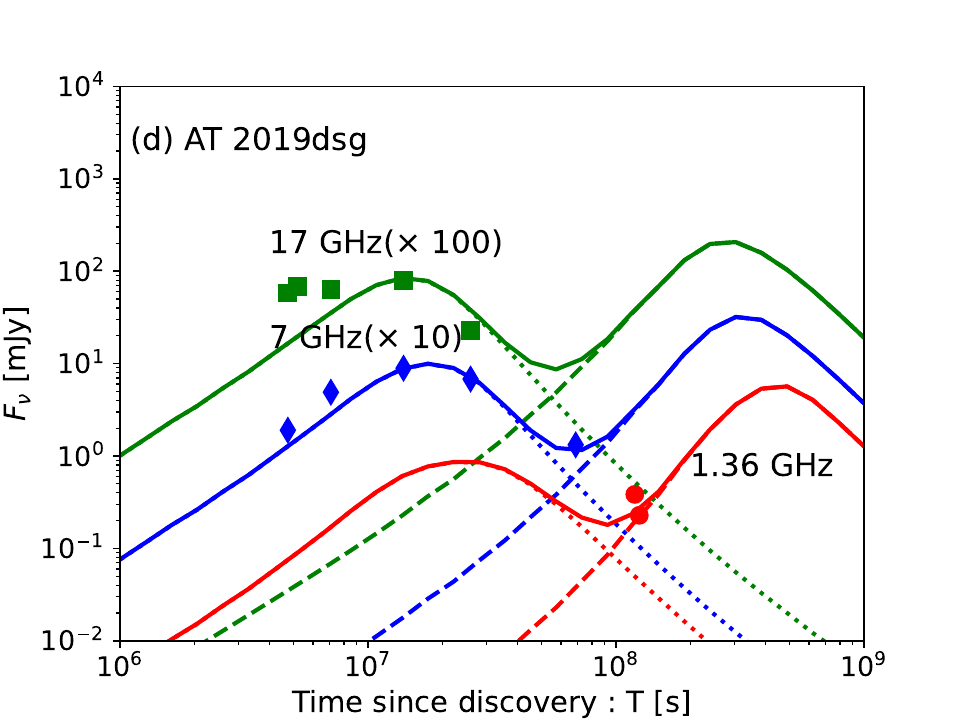}
\end{minipage}
\vspace{-0.1cm}
\caption{
Radio light curves are calculated with our two-component jet model.
The solid lines consist of the sum of emission components of the narrow (dashed lines) and wide (dotted lines) jets. 
The meanings of colours and observed data are the same as in Fig.~\ref{one-jet}. 
}
\label{two-jet}
\end{figure*}

To mimic a structured jet that is more realistic, we consider a two-component jet model, in which another ‘wide jet’ is added to the narrow jet described in Sec.~\ref{sec:one-jet}.
The observed flux is calculated by AMES as a superposition of radio emission from each jet component. 
We assume that both jet components are launched from the SMBH at the same time and in the same direction. 
The parameters of both jets are summarized in Table~\ref{tab:parameter}.
Our parameters for both jets are also consistent with the previous two-component jet scenario for on-axis TDEs~\citep{Mimica2015}.
As shown in Fig.~\ref{two-jet}, the late-time radio data at $T\gtrsim 10^8\,{\rm s}$ can be explained by the off-axis narrow jet emission (dashed lines), while the radio emission at $T\lesssim 10^8\,{\rm s}$ can be interpreted as the wide jet emission (dotted lines).

ASASSN-15oi and AT~2019dsg have the first radio peaks around $T\sim 2\times 10^7\,{\rm s}$ and at $T\sim 10^{7}\,{\rm s}$, respectively.
Our wide jet emission provides a viable explanation for the observed radio flux. The wide jet reaches the post-jet-break decay phase from $T\sim 10^6\,{\rm s}$ to $T\sim 10^9\,{\rm s}$.
When it is assumed that the wide jet enters the post-jet-break decay phase, the observer time of the flux maximum is analytically given by
\small
\begin{equation}
T_{\rm pk}\sim\cfrac{1+z}{c}\left(\cfrac{(3-w){\mathcal E}_k^{\rm iso}\theta_j^2}{4\pi \times10^{18w} n_{\rm ext} m_pc^2}\right)^{\frac{1}{3-w}}(\theta_v-\theta_j)^{2}.
\label{eq:peaktime}
\end{equation}
\normalsize
For our wide jet parameters, we obtain
$T_{\rm pk}\sim 9\times 10^{6}\,{\rm s}$ for ASASSN-14ae,
$T_{\rm pk}\sim 8\times10^{6}\,{\rm s}$ for ASASSN-15oi,
$T_{\rm pk}\sim 4\times10^{6}\,{\rm s}$ for AT~2018hyz, and
$T_{\rm pk}\sim 6\times 10^{6}\,{\rm s}$ for AT~2019dsg,
which are consistent with our numerical results within a factor of two. 
For AT~2018hyz, $\epsilon_B$ for the narrow jet is required to be larger than that for the wide jet, so the second peak is brighter. However, there is parameter degeneracy and this could be attributed to the difference in $\epsilon_e$. The situation is similar for the other 3 TDEs. For example, the second peak of AT~2019dsg can be dimmer for smaller values of $\epsilon_B$ and/or $\epsilon_e$. When the properties of CNM and/or jets are different, microphysical parameters may also be different between the narrow and wide jets \citep{Sato2023a}. More extensive data at multiwavelengths are required to better estimate the afterglow parameters. 

\section{Summary and Discussion}
\label{sec:discussion}
Radio rebrightening observed from a few TDEs on timescales of several years have been recently reported. Such a long-term rise in the radio band can be explained by either jets or delayed winds. We focused on the former scenario and showed that the two-component off-axis jet model is consistent with the radio data of ASASSN-14ae, ASASSN-15oi, AT~2018hyz, and AT~2019dsg. 

Our results for ASASSN-15oi highlight the limitation of the simplistic one-component jet model, as discussed in Section~\ref{sec:one-jet} and Fig.~\ref{one-jet}. The proposed two-component jet model with different jet opening angles is among the simplest models that can lead to two brightening episodes in the radio light curves.
Recent numerical studies have underlined the importance of such structured jets in TDEs. Rapidly spinning SMBHs with magnetically arrested  
accretion disks can naturally lead to structured jets via the Blandford-Znajek mechanism~\citep{Blandford:1977ds,McKinney:2006tf,Tchekhovskoy2011,Dai:2018jbr,Dai_2021}. The angular structure may also arise due to hydrodynamical processes~\citep{Curd:2018qvy,Lu2023}.

Better modeling of jetted TDEs is crucial for probing shock physics.
Theoretical studies have shown that particle acceleration at perpendicular relativistic shocks may not be very efficient especially for large magnetizations~\citep{Sironi:2013ri,Lemoine:2014gca,Reville:2014mta,Bell:2017zzx}. However, forward shocks considered in this work have low magnetization, and 
wide jets are typically transrelativistic, so we expect them to be efficient particle accelerators.  
Our results infer $\epsilon_B\sim{10}^{-4}-0.03$ for both narrow and wide jets, corresponding to post-shock magnetic fields of $\sim1\,{\rm G}$ at $10^7-10^8\,{\rm s}$. These inferred values of $\epsilon_B$ are consistent with those obtained from the modeling of GRB afterglows (see e.g., Refs.~\citep{Huang:2003fwa,Peng:2004gy,Sato2023a}). Various processes can be responsible for the amplification of magnetic field in the downstream (shocked CNM)~\citep{Sironi:2007as,Inoue2011,Mizuno2011,Tomita2019}. However, we note that the details of particle acceleration also depend on the upstream magnetization. The maximum energy evolves slower if the shock is Weibel-mediated~\citep{Lemoine2010,Sironi:2013ri,Sironi:2015oza}, although this may not be the case for mildly relativistic shocks~\citep{Crumley:2018kvf}. 
While we checked that the electrons can be accelerated up to sufficiently high energies to explain radio observations, we stress that further high-energy observations would be useful for probing particle acceleration (cf. Ref.~\cite{Asano:2020grw} for GRBs).

Interestingly, the radio data consistent with all TDEs may have similar values of ${\mathcal E}_k^{\rm iso}$, $\Gamma_0$, and $\theta_j$ (see table~\ref{tab:parameter} for the parameters used). To explain the rapid rise in the radio light curves, we adopt $w=0.5,\ 1.0$ as the CNM density profile index. This implies that the CNM density can be as high as $\sim{10}^2-{10}^3~{\rm cm}^{-3}$, as expected in the center of galaxies including Sgr~$\rm A^{*}$~\citep{Baganoff2003}. For AT~2019dsg, after $10^8\,{\rm s}$, the observed data can be explained by large values of $w$ for the narrow jet. However, from $T\sim4\times 10^6\,{\rm s}$ to $T\sim 1.2\times10^7\,{\rm s}$, the observed light curve at 7~GHz shows the steep rising part of $F_{\nu}\propto T^4$. Therefore, $w\sim1$ for our wide jet is favored from the early observational result at 7~GHz [see dotted lines in Fig.~\ref{two-jet}(d)]. 
While a density profile with $w=1.0$ is consistent with that of Sgr~$\rm A^{*}$ \citep{Baganoff2003}, a shallower density profile with $w=0.5$ is also reasonable for accretion flows with low viscosity \citep{Igumenshchev2000}. Moreover, other jetted scenarios \citep{Beniamini2023,Sfaradi2023} also considered the range of  $w\lesssim1$. 
We note that our model provides qualitative explanations for the radio data. The theoretical fluxes are consistent with the data only within a factor of three, and the two-component jet model would still be too simple to perform quantitative fittings. In addition to such model systematics, parameter degeneracies also exist. Despite these caveats, our modeling favors  $w\lesssim1.0-1.5$ for these 4 TDEs showing the steep rising behavior. Analytically, the radio flux in the post-jet-break phase for off-axis viewing is predicted to be $F_{\nu} \propto T^{(21-8w)/3}$. 
For $w=1.5$ that is motivated by the Bondi accretion \citep{Bondi1952} and $w=2.5$ that is used for late-time radio emission from some TDEs such as ASASSN-14li \citep{Alexander2016}, we have $F_{\nu}\propto T^{3.0}$ and $F_{\nu}\propto T^{0.3}$, respectively (see Sec.~\ref{sec:one-jet}), which are inconsistent with the observed data, especially for ASASSN-15oi and AT~2018hyz.

On-axis jetted TDEs are rarer and brighter with a radio luminosity of $\gtrsim 10^{40}~{\rm erg~s^{-1}}$. As seen from Fig.~\ref{two-jet}, off-axis narrow jets may have $\sim 10^{39}-10^{40}~{\rm erg~s^{-1}}$ around the peak that is $\sim10$~yr after the optical discovery. 
Although Ref.~\cite{Cendes2023} reported 24 TDEs that are dimmer than on-axis jetted TDEs by $2-3$ orders of magnitude, we expect that off-axis jetted TDEs compose a subdominant population. 
The apparent rate density of on-axis jetted TDEs is estimated to be $\rho_{\rm jTDE}\approx 0.03^{+0.04}_{-0.02}~{\rm Gpc^{-3}yr^{-1}}$ \citep{Sun2015}, leading to a true rate density $R_{\rm jTDE} \sim 6~{\rm Gpc^{-3} yr^{-1}}$ of jetted TDEs. 
Thus, the true rate density of jetted TDEs including both on-axis and off-axis events is expected to be a few percent of the rate density of all TDEs \citep{Sun2015}, although the all TDE rate is model dependent and has large uncertainty \citep{Sun2015,Alexander2020}.
Assuming $f_\Omega = (\theta_j-\theta_v)^2/2\sim 0.6$~sr as the typical solid angle fraction for off-axis viewing, the expected event rate within $d_L \approx 0.3$~Gpc is estimated to be
\footnotesize
\begin{eqnarray}
\dot{N}_{\rm jTDE}&\approx& \frac{4\pi}{3}d_L^3R_{\rm jTDE}f_\Omega \nonumber \\
&\sim& 0.4~{\rm yr^{-1}}\left(\frac{R_{\rm jTDE}}{6.0~{\rm Gpc^{-3}~yr^{-1}}}\right) \left(\frac{d_L}{0.3~{\rm Gpc}}\right)^3\left(\frac{f_\Omega}{0.6~{\rm sr}}\right)\,.
\label{eq:event}
\end{eqnarray}
\normalsize 
This value is smaller than the observed rate of the TDEs that exhibit delayed radio flares \citep{Cendes2023,Goodwin:2023qcq,Christy2024}. 
However, the event rate in the off-axis viewing case is consistent with radio follow-up observations of optically discovered TDEs for a typical exposure time of $\sim10^5\,{\rm s}$, a field of view of $\sim 10^{-5}-{10^{-4}}\,~{\rm rad}$, and a duty cycle of $\sim 0.1-1\%$ (see e.g., Refs.~\citep{Alexander2020,Cendes2021,Horesh2021,Cendes2022,Cendes2023}), even though the above argument is subject to observational biases \citep{Alexander2020}. 
Nevertheless, despite the large uncertainty, even with an extreme value of the viewing angle, e.g. $\theta_v\sim\pi/2$, the inferred event rate would not be enough to explain all radio-detected TDEs, which could be explained if a significant fraction of TDE jets are choked \citep{MM2023,Lu2023}. Some other TDEs especially with a radio luminosity of $\sim 10^{37}-10^{38}~{\rm erg~s^{-1}}$ may be explained by delayed disk-driven winds \citep{Cendes2023}. 

To reveal the outflow properties of radio-detected TDEs and to go beyond the one-component outflow model, 
more dedicated observations are necessary. 
First, samples of radio-detected TDEs are far from complete, and systematic surveys with existing facilities such as VLA (Very Large Array) \citep{VLA}, MeerKAT \citep{MeerKAT}, and ATCA (Australia Telescope Compact Array) \citep{ATCA}, ngVLA (Next Generation Very Large Array) \citep{ngVLA}, and SKA (Square Kilometre Array) \citep{SKA} will be useful for detecting more TDEs exhibiting late-time radio rebrightening with $\sim 10^{39}-10^{40}~{\rm erg~s^{-1}}$. 
Second, multiyear observations will be crucial for discriminating among different models and modeling the spectral and temporal evolution of radio emission from the outflows. For example, radio emission from AT~2019dsg can also be explained by the wind, so radio data at later times would be useful for testing the off-axis jet model. On the other hand, for ASASSN-14ae and AT~2018hyz, radio data at earlier times would have been beneficial.  
Third, higher-cadence observations may enable us to identify the peaks and valleys in the predicted light curves, which can also be used for constraining jet properties such as the launching time and Lorentz factor. 
Note that in this work the jet is assumed to be launched around the disruption time without any significant delay. 
However, if the disk formation is delayed, the jet launch can also be delayed \cite{MM2023}, and afterglow emission may be refreshed by late-time energy injections.  

Multiwavelength observations would be relevant for testing the jet and wind models (Sato et al., in prep). X-ray emission has been detected for some jetted TDEs \citep{Bloom2011,Burrows2011,Cenko2012,Brown2015,Andreoni2022}, which can be explained by the narrow jet viewed on-axis. 
While x-ray emission from ASASSN-14ae, ASASSN-15oi, and AT~2018hyz was not observed, AT~2019dsg exhibited x-ray emission that may come from an accretion disk and corona \citep{Cannizzaro_2021}. 
X-rays from the off-axis jet may be challenging to detect but could be seen by deep observations with {\it XMM-Newton} (X‐ray Multi-Mirror Mission ‐ Newton) \citep{XMM-Newton} and the {\it Chandra} X-ray Observatory \citep{Chandra} for nearby TDEs and/or with next-generation x-ray telescopes such as {\it Athena} (Advanced Telescope for the High ENergy Astrophysics) \citep{Athena} and {\it eROSITA} (extended Roentgen Survey with an Imaging Telescope Array) \citep{eROSITA}.
Moreover, quasi-simultaneous optical observations with, e.g., the Vera C. Rubin Observatory \citep{Rubin} will be useful for testing the models. 
Multimessenger observations may also provide additional information. Recently, coincident high-energy neutrino events have been reported for several TDEs including AT~2019dsg \citep{Stein2021}, AT~2019fdr \citep{Reusch2022}, and AT~2019aalc \citep{vVelzen2021}) with a possible time delay of $\sim 150-400\,{\rm days}$ after their optical discoveries. 
The on-axis jet model is unlikely for AT~2019dsg  \citep{Murase:2020lnu,Cendes2021,Matsumoto:2021qqo}, and neutrino production in disks \cite{Hayasaki2019,Murase:2020lnu}, coronae \cite{Murase:2020lnu}, winds \cite{Murase:2020lnu,Winter:2020ptf,Yuan:2024foi}, and choked jets \cite{Wang:2015mmh,Senno:2016bso,MM2023} have been considered. 

\

\section*{Acknowledgements}
We thank 
Kimitake~Hayasaki, 
Wenbin~Lu,
Tatsuya~Matsumoto,
Rin~Oikawa, 
Shuta~J.~Tanaka, and
Ryo~Yamazaki
for valuable comments.
We also thank an anonymous referee for their useful comments which have improved this paper.
This research was partially supported by JSPS KAKENHI, grant numbers 22KJ2643 (YS), 20H01901 and 20H05852 (KM).
The work of K.M. is supported by the NSF grants Nos. AST-2108466, AST-2108467, and AST-2308021.
M.B. acknowledges support from the Eberly Postdoctoral Research Fellowship at the Pennsylvania State University.
J.C. is supported by the NSF Grant Nos. AST-1908689 and AST-2108466.
J.C. also acknowledges the Nevada Center for Astrophysics and NASA award 80NSSC23M0104 for support.
M.M. also acknowledges support from the Institute for Gravitation and the Cosmos (IGC) Postdoctoral Fellowship.

\

\section*{Appendix A: SUPPLEMENTARY MATERIAL}

Astrophysical Multimessenger Emission Simulator ({\sc AMES}) allows us to compute synchrotron and inverse-Compton emission from outflows, where synchrotron self-absorption and two-photon annihilation attenuation are included. 
Its blastwave afterglow module, which can be used for gamma-ray bursts (GRBs), magnetar flares, and tidal disruption events (TDEs), considers external forward and reverse shocks with/without energy injections~\cite{Zhang2021,Wei:2023rmr,Zhang:2023uei}. 
The single-zone model can be used as an approximation for on-axis afterglow emission, but it is essential to consider the equal-arrival-time-surface (EATS) to consider off-axis afterglow emission. Here we describe treatments on afterglow dynamics and EATS calculations implemented in {\sc AMES}.

To describe the evolution of the bulk Lorentz factor, $\Gamma$, we solve the following differential equation~\cite[e.g.,][]{Nava:2012hq},
\begin{align}
    \frac{d\Gamma}{dR} &= -\frac{(\Gamma_{\rm eff} + 1)(\Gamma - 1) c^2 \frac{dm}{dR} + \Gamma_{\rm eff} \frac{d\mathcal{E}_{\rm ad}^\prime}{dR}}{ (M_{\rm ej} + m) c^2 + \mathcal{E}_{\rm int}^\prime \frac{d\Gamma_{\rm eff}}{d\Gamma}},
\end{align}
where $\Gamma_{\rm eff} \equiv (\hat{\gamma}\Gamma^2 - \hat{\gamma} + 1)/\Gamma$, $\hat{\gamma}$ is the adiabatic index,
\begin{equation}
    M_{\rm ej} = \frac{\mathcal{E}_k^{\rm iso} (1 - {\rm cos}\theta_j)}{2\Gamma_0 c^2},
\end{equation}
is the total mass of the ejecta and
\begin{equation}
    dm = 2\pi (1 - {\rm cos}\theta_j) R^2 n_{\rm ext} m_H dR,
\end{equation}
is the total mass of the swept-up matter within $dR$. Here $n_{\rm ext}$ is the external density at $R$, $\theta_j$ is the jet opening angle, and $m_H$ is the hydrogen mass, where the proton-electron plasma is assumed. The evolution of internal energy and energy lost due to the adiabatic expansion is~\cite[e.g.,][]{zhangPhysicsGammaRayBursts2018}
\begin{equation}\label{eq:int}
    \frac{d\mathcal{E}_{\rm int}^\prime}{dR} = (1 - \epsilon) (\Gamma - 1) 4\pi R^2 n_{\rm ext} m_p c^2 + \frac{d\mathcal{E}_{\rm ad}^\prime}{dR},
\end{equation}
and
\begin{equation}\label{eq:adiabatic}
    \frac{d\mathcal{E}_{\rm ad}^\prime}{dR} = -(\hat{\gamma} - 1) \left(\frac{3}{R} - \frac{1}{\Gamma} \frac{d\Gamma}{dR}\right) \mathcal{E}_{\rm int}^\prime.
\end{equation}
The blastwave afterglow module of {\sc AMES} solves Eq.~(S1) and utilizes Eq.~(1) in the main text, and this work focuses on the thin shell region. However, as described in Ref.~\cite{Zhang:2023uei}, it is possible to consider the thick shell regime, and not only the forward shock but also the reverse shock can be taken into account (see Appendix of Ref.~\cite{Zhang:2023uei}). Note that Ref.~\cite{Zhang2021} ignores the term of adiabatic energy losses, and uses the expression for on-axis emission (see their Eq.~C2) or the single-zone approximation~\cite{Murase:2010fq}. 
In this work, we ignore the lateral expansion although we do take into account the jet break due to the debeaming. However, for the tophat jet, it is possible to consider the simplest lateral expansion~\cite{zhangPhysicsGammaRayBursts2018}.  

\begin{figure}[t]
    \centering
    \includegraphics[width=0.9\linewidth]{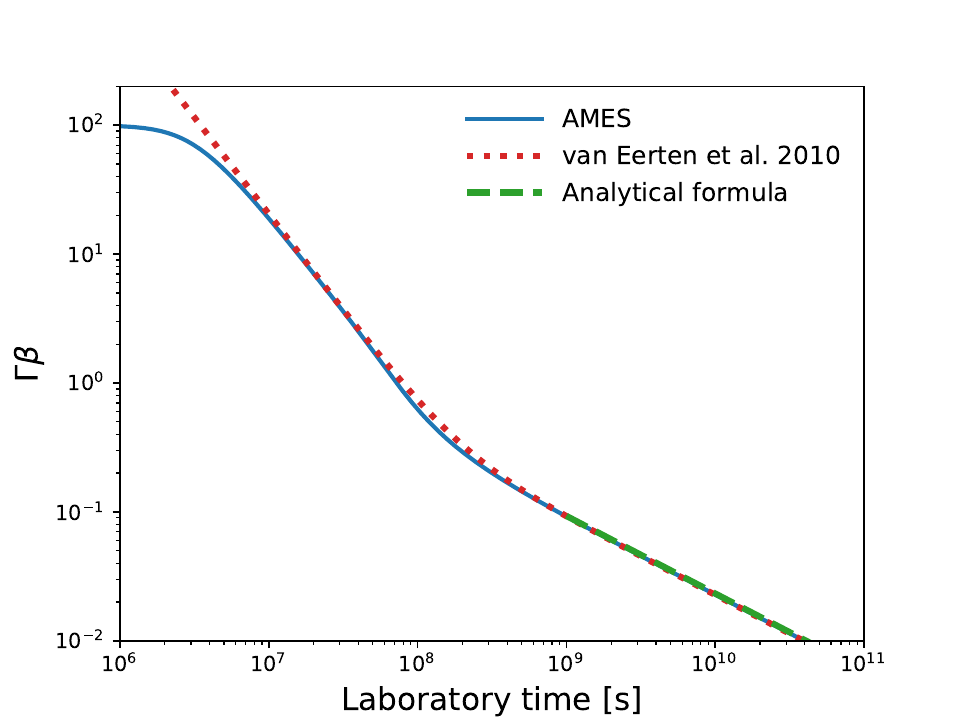}
        \includegraphics[width=0.9\linewidth]{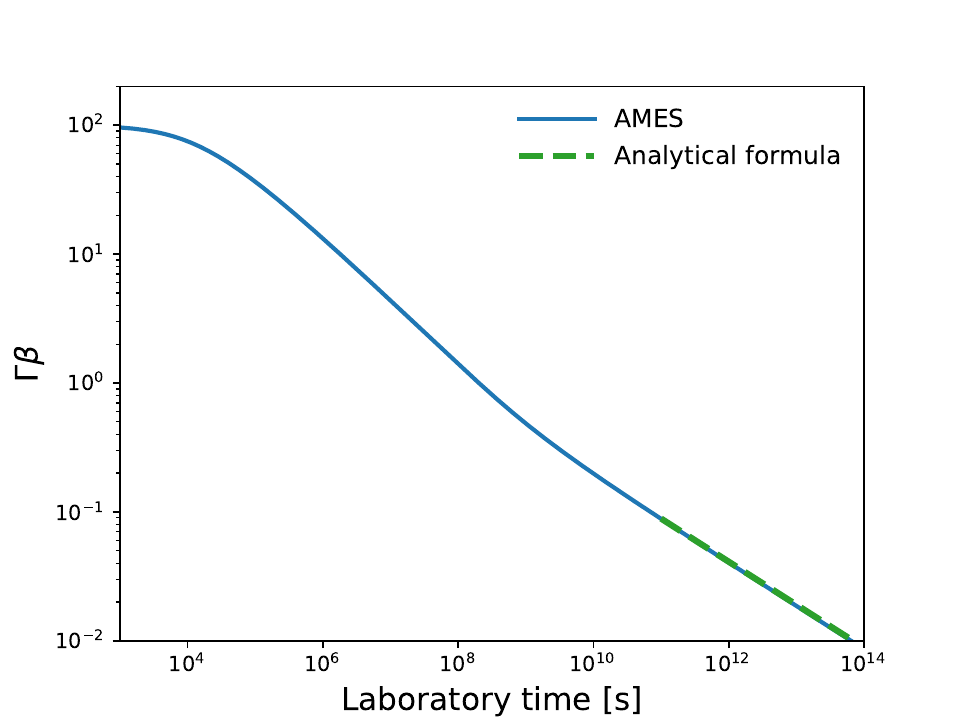}
    \caption{Upper panel: Dynamical evolution of $\Gamma \beta$ as a function of the laboratory time in the constant density with $n_{\rm ext} = 1\rm~cm^{-3}$. 
    Lower panel: Same as the upper panel but for the wind density profile with $n_{\rm ext} = 1\rm~cm^{-3}$, $r_{\rm ext} = 10^{18}\rm~cm$, and $w = 2$. In both panels, the used parameters are $\Gamma_0 = 100$ and $\mathcal{E}_k^{\rm iso} =5\times10^{52}\rm~erg$.}
    \label{fig:dynamic}
\end{figure}

\begin{figure}[t]
    \centering
    \includegraphics[width=1.0\linewidth]{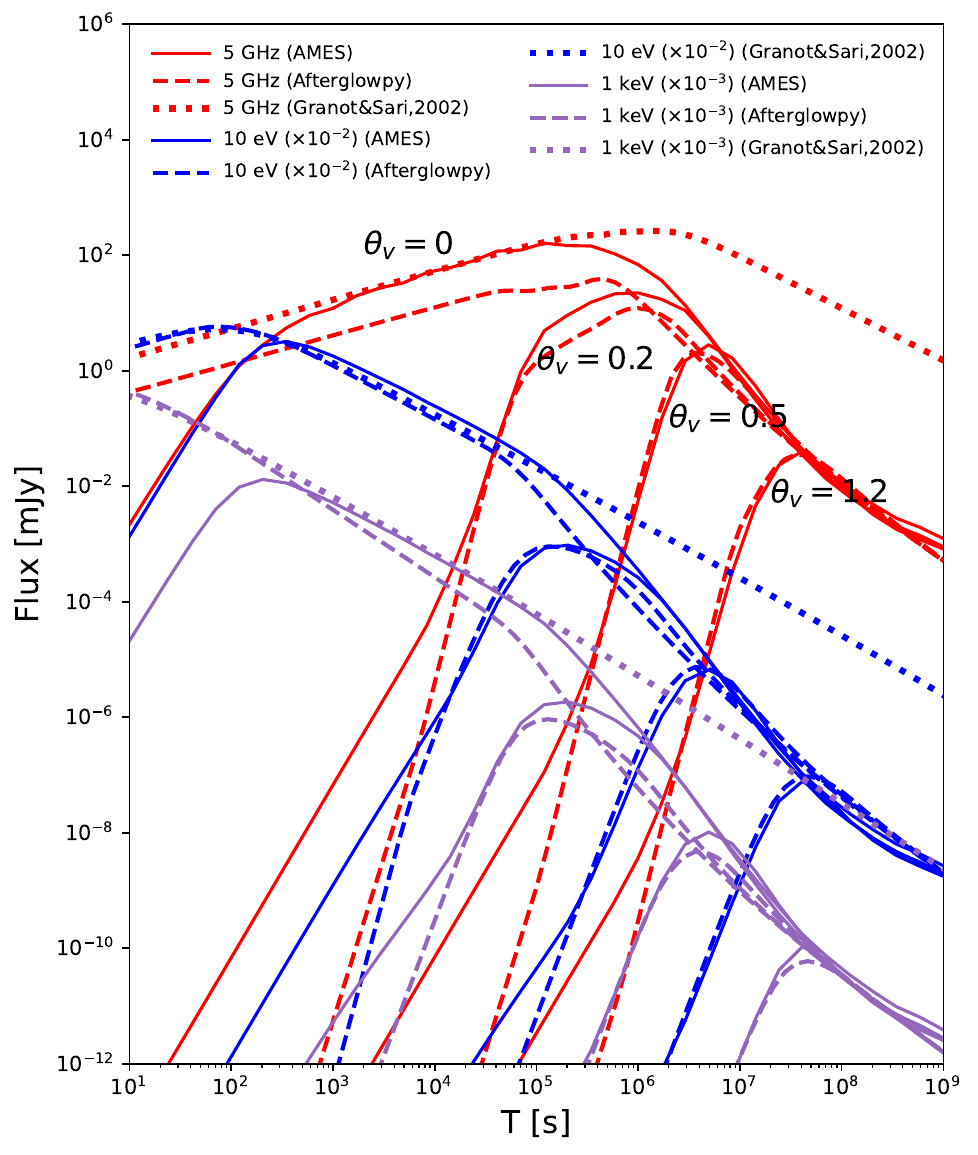}
    \caption{On-axis and off-axis light curves of forward shock afterglow emission calculated with {\sc AMES} (solid curves) and {\sc Afterglowpy} (dashed curves), respectively. The parameters are $\Gamma_0 = 100$, $\mathcal{E}_k^{\rm iso} = 5\times 10^{52}\rm~erg$, $n_{\rm ext} = 1\rm~cm^{-3}$ ($w=0$), $s = 2.2$, $\epsilon_e = 0.1$, $\epsilon_B = 0.001$, $f_e=1$, and $\theta_j = 0.1$.}
    \label{fig:lightcurve}
\end{figure}

Our model corresponds to the homogeneous shell model, and the exact self-similar solutions give different normalizations. To recover the Blandford-McKee solution in the relativistic limit~\cite{Blandford:1976uq}, following Ref.~\cite{Nava:2012hq}, we use
\begin{equation}
C_{\rm BM}= \epsilon + \frac{9- 2 w}{17 - 4w} (1 - \epsilon),  
\end{equation}
where $\epsilon = 0$ in the adiabatic case and $\epsilon = 1$ in the radiative case. We focus on the adiabatic regime, and for the Sedov-Taylor solution in the nonrelativistic limit~\cite{Sedov59,Taylor50,Ostriker:1988zz} we adopt
\begin{equation}
C_{\rm ST}=\frac{16}{9} \frac{(3-w)(5-w)}{10-3w}. 
\end{equation} 
To calculate the transition regime, we introduce
\begin{equation}
C=\beta(C_{\rm BM}-C_{\rm ST})+C_{\rm ST}, 
\end{equation}
where $\beta=\sqrt{1 - 1 / \Gamma^2}$.
We multiply $n_{\rm ext}$ by this correction factor in Eqs.~(\ref{eq:int}) and (\ref{eq:adiabatic}). 

As shown in Fig.~\ref{fig:dynamic}, the above interpolation formula reproduces both relativistic and nonrelativistic regimes consistently. 
When the external density profile is constant, we compare our results to Eq.~(A12) in Ref.~\cite{vanEerten:2010zh} and the analytical formula of the Sedov-Taylor solution, considering that the shock velocity is $\beta_{\rm sh}\approx(4/3)\beta$~\cite{Sedov59,Taylor50,Ostriker:1988zz,Wei:2023rmr}. 
For the wind density profile, we compare our results to Eq.~(79) in Ref.~\cite{Piro:2018jpl}. In both cases, the numerical results show good agreement with the analytical formulas.

Performing the integral of Eq.~(1) in the main text leads to light curves and spectra of afterglow emission from the forward shock. Note that the Lorentz factor of the shock is given by~\cite{Blandford:1976uq} 
\begin{equation}
\Gamma_{\rm sh}^2 =\frac{1}{1-\beta_{\rm sh}^2}= \frac{(\Gamma + 1)[\hat{\gamma}(\Gamma - 1) + 1]^2}{\hat{\gamma}(2 - \hat{\gamma})(\Gamma - 1)+2},
\end{equation}
where $\Gamma$ is the Lorentz factor of the shocked shell. For highly relativistic shocks, we have $\Gamma_{\rm sh} \approx \sqrt{2}\Gamma$, while we expect $\beta_{\rm sh}\approx(4/3)\beta$ in the nonrelativistic limit.

Example light curves are shown in Fig.~\ref{fig:lightcurve}. 
The light curves of {\rm AMES} are consistent with the results of Ref.~\cite{Granot:2001ge} before the jet break. For comparison, the light curves of {\sc afterglowpy} are also shown. Note that {\sc AMES} takes into account synchrotron self-absorption in the radio band and radiative cooling due to inverse-Compton radiation. The difference in early time light curves comes from the coasting phase, and the flattening of the light curves at very late times is consistent with the predicted behavior in the deep Newtonian phase~\cite{Wei:2023rmr}.

\bibliography{TDE}

\begin{thebibliography}{117}%
\makeatletter
\providecommand \@ifxundefined [1]{%
 \@ifx{#1\undefined}
}%
\providecommand \@ifnum [1]{%
 \ifnum #1\expandafter \@firstoftwo
 \else \expandafter \@secondoftwo
 \fi
}%
\providecommand \@ifx [1]{%
 \ifx #1\expandafter \@firstoftwo
 \else \expandafter \@secondoftwo
 \fi
}%
\providecommand \natexlab [1]{#1}%
\providecommand \enquote  [1]{``#1''}%
\providecommand \bibnamefont  [1]{#1}%
\providecommand \bibfnamefont [1]{#1}%
\providecommand \citenamefont [1]{#1}%
\providecommand \href@noop [0]{\@secondoftwo}%
\providecommand \href [0]{\begingroup \@sanitize@url \@href}%
\providecommand \@href[1]{\@@startlink{#1}\@@href}%
\providecommand \@@href[1]{\endgroup#1\@@endlink}%
\providecommand \@sanitize@url [0]{\catcode `\\12\catcode `\$12\catcode
  `\&12\catcode `\#12\catcode `\^12\catcode `\_12\catcode `\%12\relax}%
\providecommand \@@startlink[1]{}%
\providecommand \@@endlink[0]{}%
\providecommand \url  [0]{\begingroup\@sanitize@url \@url }%
\providecommand \@url [1]{\endgroup\@href {#1}{\urlprefix }}%
\providecommand \urlprefix  [0]{URL }%
\providecommand \Eprint [0]{\href }%
\providecommand \doibase [0]{https://doi.org/}%
\providecommand \selectlanguage [0]{\@gobble}%
\providecommand \bibinfo  [0]{\@secondoftwo}%
\providecommand \bibfield  [0]{\@secondoftwo}%
\providecommand \translation [1]{[#1]}%
\providecommand \BibitemOpen [0]{}%
\providecommand \bibitemStop [0]{}%
\providecommand \bibitemNoStop [0]{.\EOS\space}%
\providecommand \EOS [0]{\spacefactor3000\relax}%
\providecommand \BibitemShut  [1]{\csname bibitem#1\endcsname}%
\let\auto@bib@innerbib\@empty
%</preamble>
\bibitem [{\citenamefont {{Hills}}(1975)}]{Hills1975}%
  \BibitemOpen
  \bibfield  {author} {\bibinfo {author} {\bibfnamefont {J.~G.}\ \bibnamefont
  {{Hills}}},\ }\bibfield  {title} {\bibinfo {title} {{Possible power source of
  Seyfert galaxies and QSOs}},\ }\href {https://doi.org/10.1038/254295a0}
  {\bibfield  {journal} {\bibinfo  {journal} {\nat}\ }\textbf {\bibinfo
  {volume} {254}},\ \bibinfo {pages} {295} (\bibinfo {year}
  {1975})}\BibitemShut {NoStop}%
\bibitem [{\citenamefont {Rees}(1988)}]{Rees1988}%
  \BibitemOpen
  \bibfield  {author} {\bibinfo {author} {\bibfnamefont {M.~J.}\ \bibnamefont
  {Rees}},\ }\bibfield  {title} {\bibinfo {title} {{Tidal disruption of stars
  by black holes of 10 to the 6th-10 to the 8th solar masses in nearby
  galaxies}},\ }\href {https://doi.org/10.1038/333523a0} {\bibfield  {journal}
  {\bibinfo  {journal} {Nature}\ }\textbf {\bibinfo {volume} {333}},\ \bibinfo
  {pages} {523} (\bibinfo {year} {1988})}\BibitemShut {NoStop}%
\bibitem [{\citenamefont {{Evans}}\ and\ \citenamefont
  {{Kochanek}}(1989)}]{EK1989}%
  \BibitemOpen
  \bibfield  {author} {\bibinfo {author} {\bibfnamefont {C.~R.}\ \bibnamefont
  {{Evans}}}\ and\ \bibinfo {author} {\bibfnamefont {C.~S.}\ \bibnamefont
  {{Kochanek}}},\ }\bibfield  {title} {\bibinfo {title} {{The Tidal Disruption
  of a Star by a Massive Black Hole}},\ }\href {https://doi.org/10.1086/185567}
  {\bibfield  {journal} {\bibinfo  {journal} {\apjl}\ }\textbf {\bibinfo
  {volume} {346}},\ \bibinfo {pages} {L13} (\bibinfo {year}
  {1989})}\BibitemShut {NoStop}%
\bibitem [{\citenamefont {Loeb}\ and\ \citenamefont {Ulmer}(1997)}]{Loeb1997}%
  \BibitemOpen
  \bibfield  {author} {\bibinfo {author} {\bibfnamefont {A.}~\bibnamefont
  {Loeb}}\ and\ \bibinfo {author} {\bibfnamefont {A.}~\bibnamefont {Ulmer}},\
  }\bibfield  {title} {\bibinfo {title} {{Optical appearance of the debris of a
  star disrupted by a massive black hole}},\ }\href
  {https://doi.org/10.1086/304814} {\bibfield  {journal} {\bibinfo  {journal}
  {Astrophys. J.}\ }\textbf {\bibinfo {volume} {489}},\ \bibinfo {pages} {573}
  (\bibinfo {year} {1997})}\BibitemShut {NoStop}%
\bibitem [{\citenamefont {Strubbe}\ and\ \citenamefont
  {Quataert}(2009)}]{Strubbe2009}%
  \BibitemOpen
  \bibfield  {author} {\bibinfo {author} {\bibfnamefont {L.~E.}\ \bibnamefont
  {Strubbe}}\ and\ \bibinfo {author} {\bibfnamefont {E.}~\bibnamefont
  {Quataert}},\ }\bibfield  {title} {\bibinfo {title} {{Optical Flares from the
  Tidal Disruption of Stars by Massive Black Holes}},\ }\href
  {https://doi.org/10.1111/j.1365-2966.2009.15599.x} {\bibfield  {journal}
  {\bibinfo  {journal} {Mon. Not. Roy. Astron. Soc.}\ }\textbf {\bibinfo
  {volume} {400}},\ \bibinfo {pages} {2070} (\bibinfo {year}
  {2009})}\BibitemShut {NoStop}%
\bibitem [{\citenamefont {Lodato}\ and\ \citenamefont
  {Rossi}(2011)}]{Lodato2011}%
  \BibitemOpen
  \bibfield  {author} {\bibinfo {author} {\bibfnamefont {G.}~\bibnamefont
  {Lodato}}\ and\ \bibinfo {author} {\bibfnamefont {E.}~\bibnamefont {Rossi}},\
  }\bibfield  {title} {\bibinfo {title} {{Multiband lightcurves of tidal
  disruption events}},\ }\href
  {https://doi.org/10.1111/j.1365-2966.2010.17448.x} {\bibfield  {journal}
  {\bibinfo  {journal} {Mon. Not. Roy. Astron. Soc.}\ }\textbf {\bibinfo
  {volume} {410}},\ \bibinfo {pages} {359} (\bibinfo {year}
  {2011})}\BibitemShut {NoStop}%
\bibitem [{\citenamefont {Komossa}(2015)}]{Komossa2015}%
  \BibitemOpen
  \bibfield  {author} {\bibinfo {author} {\bibfnamefont {S.}~\bibnamefont
  {Komossa}},\ }\bibfield  {title} {\bibinfo {title} {{Tidal disruption of
  stars by supermassive black holes: Status of observations}},\ }\href
  {https://doi.org/10.1016/j.jheap.2015.04.006} {\bibfield  {journal} {\bibinfo
   {journal} {JHEAp}\ }\textbf {\bibinfo {volume} {7}},\ \bibinfo {pages} {148}
  (\bibinfo {year} {2015})}\BibitemShut {NoStop}%
\bibitem [{\citenamefont {Metzger}\ and\ \citenamefont {Stone}(2016)}]{MS2016}%
  \BibitemOpen
  \bibfield  {author} {\bibinfo {author} {\bibfnamefont {B.~D.}\ \bibnamefont
  {Metzger}}\ and\ \bibinfo {author} {\bibfnamefont {N.~C.}\ \bibnamefont
  {Stone}},\ }\bibfield  {title} {\bibinfo {title} {{A bright year for tidal
  disruptions}},\ }\href {https://doi.org/10.1093/mnras/stw1394} {\bibfield
  {journal} {\bibinfo  {journal} {Mon. Not. Roy. Astron. Soc.}\ }\textbf
  {\bibinfo {volume} {461}},\ \bibinfo {pages} {948} (\bibinfo {year}
  {2016})}\BibitemShut {NoStop}%
\bibitem [{\citenamefont {{Roth}}\ \emph {et~al.}(2016)\citenamefont {{Roth}}
  \emph {et~al.}}]{Roth2016}%
  \BibitemOpen
  \bibfield  {author} {\bibinfo {author} {\bibfnamefont {N.}~\bibnamefont
  {{Roth}}} \emph {et~al.},\ }\bibfield  {title} {\bibinfo {title} {{The X-Ray
  through Optical Fluxes and Line Strengths of Tidal Disruption Events}},\
  }\href {https://doi.org/10.3847/0004-637X/827/1/3} {\bibfield  {journal}
  {\bibinfo  {journal} {\apj}\ }\textbf {\bibinfo {volume} {827}},\ \bibinfo
  {eid} {3} (\bibinfo {year} {2016})}\BibitemShut {NoStop}%
\bibitem [{\citenamefont {De~Colle}\ \emph {et~al.}(2012)\citenamefont
  {De~Colle} \emph {et~al.}}]{DeColle2012}%
  \BibitemOpen
  \bibfield  {author} {\bibinfo {author} {\bibfnamefont {F.}~\bibnamefont
  {De~Colle}} \emph {et~al.},\ }\bibfield  {title} {\bibinfo {title} {{The
  dynamics, appearance and demographics of relativistic jets triggered by tidal
  disruption of stars in quiescent supermassive black holes}},\ }\href
  {https://doi.org/10.1088/0004-637X/760/2/103} {\bibfield  {journal} {\bibinfo
   {journal} {Astrophys. J.}\ }\textbf {\bibinfo {volume} {760}},\ \bibinfo
  {pages} {103} (\bibinfo {year} {2012})}\BibitemShut {NoStop}%
\bibitem [{\citenamefont {{Alexander}}\ \emph {et~al.}(2020)\citenamefont
  {{Alexander}} \emph {et~al.}}]{Alexander2020}%
  \BibitemOpen
  \bibfield  {author} {\bibinfo {author} {\bibfnamefont {K.~D.}\ \bibnamefont
  {{Alexander}}} \emph {et~al.},\ }\bibfield  {title} {\bibinfo {title} {{Radio
  Properties of Tidal Disruption Events}},\ }\href
  {https://doi.org/10.1007/s11214-020-00702-w} {\bibfield  {journal} {\bibinfo
  {journal} {Space Sci. Rev.}\ }\textbf {\bibinfo {volume} {216}},\ \bibinfo
  {pages} {81} (\bibinfo {year} {2020})}\BibitemShut {NoStop}%
\bibitem [{\citenamefont {van Velzen}\ \emph {et~al.}(2021)\citenamefont {van
  Velzen} \emph {et~al.}}]{van_Velzen2021}%
  \BibitemOpen
  \bibfield  {author} {\bibinfo {author} {\bibfnamefont {S.}~\bibnamefont {van
  Velzen}} \emph {et~al.},\ }\bibfield  {title} {\bibinfo {title} {{Seventeen
  Tidal Disruption Events from the First Half of ZTF Survey Observations:
  Entering a New Era of Population Studies}},\ }\href
  {https://doi.org/10.3847/1538-4357/abc258} {\bibfield  {journal} {\bibinfo
  {journal} {Astrophys. J.}\ }\textbf {\bibinfo {volume} {908}},\ \bibinfo
  {pages} {4} (\bibinfo {year} {2021})}\BibitemShut {NoStop}%
\bibitem [{\citenamefont {Bloom}\ \emph {et~al.}(2011)\citenamefont {Bloom}
  \emph {et~al.}}]{Bloom2011}%
  \BibitemOpen
  \bibfield  {author} {\bibinfo {author} {\bibfnamefont {J.~S.}\ \bibnamefont
  {Bloom}} \emph {et~al.},\ }\bibfield  {title} {\bibinfo {title} {{A
  relativistic jetted outburst from a massive black hole fed by a tidally
  disrupted star}},\ }\href {https://doi.org/10.1126/science.1207150}
  {\bibfield  {journal} {\bibinfo  {journal} {Science}\ }\textbf {\bibinfo
  {volume} {333}},\ \bibinfo {pages} {203} (\bibinfo {year}
  {2011})}\BibitemShut {NoStop}%
\bibitem [{\citenamefont {Burrows}\ \emph {et~al.}(2011)\citenamefont {Burrows}
  \emph {et~al.}}]{Burrows2011}%
  \BibitemOpen
  \bibfield  {author} {\bibinfo {author} {\bibfnamefont {D.~N.}\ \bibnamefont
  {Burrows}} \emph {et~al.},\ }\bibfield  {title} {\bibinfo {title} {{Discovery
  of the Onset of Rapid Accretion by a Dormant Massive Black Hole}},\ }\href
  {https://doi.org/10.1038/nature10374} {\bibfield  {journal} {\bibinfo
  {journal} {Nature}\ }\textbf {\bibinfo {volume} {476}},\ \bibinfo {pages}
  {421} (\bibinfo {year} {2011})}\BibitemShut {NoStop}%
\bibitem [{\citenamefont {Cenko}\ \emph {et~al.}(2012)\citenamefont {Cenko}
  \emph {et~al.}}]{Cenko2012}%
  \BibitemOpen
  \bibfield  {author} {\bibinfo {author} {\bibfnamefont {S.~B.}\ \bibnamefont
  {Cenko}} \emph {et~al.},\ }\bibfield  {title} {\bibinfo {title} {{Swift
  J2058.4+0516: Discovery of a Possible Second Relativistic Tidal Disruption
  Flare}},\ }\href {https://doi.org/10.1088/0004-637X/753/1/77} {\bibfield
  {journal} {\bibinfo  {journal} {Astrophys. J.}\ }\textbf {\bibinfo {volume}
  {753}},\ \bibinfo {pages} {77} (\bibinfo {year} {2012})}\BibitemShut
  {NoStop}%
\bibitem [{\citenamefont {Brown}\ \emph {et~al.}(2015)\citenamefont {Brown}
  \emph {et~al.}}]{Brown2015}%
  \BibitemOpen
  \bibfield  {author} {\bibinfo {author} {\bibfnamefont {G.~C.}\ \bibnamefont
  {Brown}} \emph {et~al.},\ }\bibfield  {title} {\bibinfo {title} {{Swift
  J1112.2\ensuremath{-}8238: a candidate relativistic tidal disruption
  flare}},\ }\href {https://doi.org/10.1093/mnras/stv1520} {\bibfield
  {journal} {\bibinfo  {journal} {\mnras}\ }\textbf {\bibinfo {volume} {452}},\
  \bibinfo {pages} {4297} (\bibinfo {year} {2015})}\BibitemShut {NoStop}%
\bibitem [{\citenamefont {Andreoni}\ \emph {et~al.}(2022)\citenamefont
  {Andreoni} \emph {et~al.}}]{Andreoni2022}%
  \BibitemOpen
  \bibfield  {author} {\bibinfo {author} {\bibfnamefont {I.}~\bibnamefont
  {Andreoni}} \emph {et~al.},\ }\bibfield  {title} {\bibinfo {title} {{A very
  luminous jet from the disruption of a star by a massive black hole}},\ }\href
  {https://doi.org/10.1038/s41586-022-05465-8} {\bibfield  {journal} {\bibinfo
  {journal} {\nat}\ }\textbf {\bibinfo {volume} {612}},\ \bibinfo {pages} {430}
  (\bibinfo {year} {2022})},\ \bibinfo {note} {[Erratum: Nature 613, E6
  (2023)]}\BibitemShut {NoStop}%
\bibitem [{\citenamefont {{Teboul}}\ \emph {et~al.}(2023)\citenamefont
  {{Teboul}}, \citenamefont {{Stone}},\ and\ \citenamefont
  {{Ostriker}}}]{Teboul2022}%
  \BibitemOpen
  \bibfield  {author} {\bibinfo {author} {\bibfnamefont {O.}~\bibnamefont
  {{Teboul}}}, \bibinfo {author} {\bibfnamefont {N.~C.}\ \bibnamefont
  {{Stone}}},\ and\ \bibinfo {author} {\bibfnamefont {J.~P.}\ \bibnamefont
  {{Ostriker}}},\ }\bibfield  {title} {\bibinfo {title} {{Loss cone
  shielding}},\ }\href {https://doi.org/10.1093/mnras/stad3301} {\bibfield
  {journal} {\bibinfo  {journal} {Mon. Not. Roy. Astron. Soc.}\ }\textbf
  {\bibinfo {volume} {527}},\ \bibinfo {pages} {3094} (\bibinfo {year}
  {2023})}\BibitemShut {NoStop}%
\bibitem [{\citenamefont {Holoien}\ \emph {et~al.}(2014)\citenamefont {Holoien}
  \emph {et~al.}}]{Holoien2014}%
  \BibitemOpen
  \bibfield  {author} {\bibinfo {author} {\bibfnamefont {T.~W.~S.}\
  \bibnamefont {Holoien}} \emph {et~al.},\ }\bibfield  {title} {\bibinfo
  {title} {{ASASSN-14ae: A Tidal Disruption Event at 200 Mpc}},\ }\href
  {https://doi.org/10.1093/mnras/stu1922} {\bibfield  {journal} {\bibinfo
  {journal} {Mon. Not. Roy. Astron. Soc.}\ }\textbf {\bibinfo {volume} {445}},\
  \bibinfo {pages} {3263} (\bibinfo {year} {2014})}\BibitemShut {NoStop}%
\bibitem [{\citenamefont {Cendes}\ \emph {et~al.}(2022)\citenamefont {Cendes}
  \emph {et~al.}}]{Cendes2022}%
  \BibitemOpen
  \bibfield  {author} {\bibinfo {author} {\bibfnamefont {Y.}~\bibnamefont
  {Cendes}} \emph {et~al.},\ }\bibfield  {title} {\bibinfo {title} {{A Mildly
  Relativistic Outflow Launched Two Years after Disruption in Tidal Disruption
  Event AT2018hyz}},\ }\href {https://doi.org/10.3847/1538-4357/ac88d0}
  {\bibfield  {journal} {\bibinfo  {journal} {Astrophys. J.}\ }\textbf
  {\bibinfo {volume} {938}},\ \bibinfo {pages} {28} (\bibinfo {year}
  {2022})}\BibitemShut {NoStop}%
\bibitem [{\citenamefont {{Goodwin}}\ \emph {et~al.}(2022)\citenamefont
  {{Goodwin}} \emph {et~al.}}]{Goodwin2022}%
  \BibitemOpen
  \bibfield  {author} {\bibinfo {author} {\bibfnamefont {A.~J.}\ \bibnamefont
  {{Goodwin}}} \emph {et~al.},\ }\bibfield  {title} {\bibinfo {title}
  {{AT2019azh: an unusually long-lived, radio-bright thermal tidal disruption
  event}},\ }\href {https://doi.org/10.1093/mnras/stac333} {\bibfield
  {journal} {\bibinfo  {journal} {Mon. Not. Roy. Astron. Soc.}\ }\textbf
  {\bibinfo {volume} {511}},\ \bibinfo {pages} {5328} (\bibinfo {year}
  {2022})}\BibitemShut {NoStop}%
\bibitem [{\citenamefont {{Horesh}}\ \emph {et~al.}(2021)\citenamefont
  {{Horesh}} \emph {et~al.}}]{Horesh2021b}%
  \BibitemOpen
  \bibfield  {author} {\bibinfo {author} {\bibfnamefont {A.}~\bibnamefont
  {{Horesh}}} \emph {et~al.},\ }\bibfield  {title} {\bibinfo {title} {{Are
  Delayed Radio Flares Common in Tidal Disruption Events? The Case of the TDE
  iPTF 16fnl}},\ }\href {https://doi.org/10.3847/2041-8213/ac25fe} {\bibfield
  {journal} {\bibinfo  {journal} {Astrophys. J. Lett.}\ }\textbf {\bibinfo
  {volume} {920}},\ \bibinfo {pages} {L5} (\bibinfo {year} {2021})}\BibitemShut
  {NoStop}%
\bibitem [{\citenamefont {{Stein}}\ \emph {et~al.}(2021)\citenamefont {{Stein}}
  \emph {et~al.}}]{Stein2021}%
  \BibitemOpen
  \bibfield  {author} {\bibinfo {author} {\bibfnamefont {R.}~\bibnamefont
  {{Stein}}} \emph {et~al.},\ }\bibfield  {title} {\bibinfo {title} {{A tidal
  disruption event coincident with a high-energy neutrino}},\ }\href
  {https://doi.org/10.1038/s41550-020-01295-8} {\bibfield  {journal} {\bibinfo
  {journal} {Nature Astron.}\ }\textbf {\bibinfo {volume} {5}},\ \bibinfo
  {pages} {510} (\bibinfo {year} {2021})}\BibitemShut {NoStop}%
\bibitem [{\citenamefont {Horesh}\ \emph {et~al.}(2021)\citenamefont {Horesh},
  \citenamefont {Cenko},\ and\ \citenamefont {Arcavi}}]{Horesh2021}%
  \BibitemOpen
  \bibfield  {author} {\bibinfo {author} {\bibfnamefont {A.}~\bibnamefont
  {Horesh}}, \bibinfo {author} {\bibfnamefont {S.~B.}\ \bibnamefont {Cenko}},\
  and\ \bibinfo {author} {\bibfnamefont {I.}~\bibnamefont {Arcavi}},\
  }\bibfield  {title} {\bibinfo {title} {{Delayed Radio Flares from a Tidal
  Disruption Event}},\ }\href {https://doi.org/10.1038/s41550-021-01300-8}
  {\bibfield  {journal} {\bibinfo  {journal} {Nature Astron.}\ }\textbf
  {\bibinfo {volume} {5}},\ \bibinfo {pages} {491} (\bibinfo {year}
  {2021})}\BibitemShut {NoStop}%
\bibitem [{\citenamefont {{Cendes}}\ \emph {et~al.}(2023)\citenamefont
  {{Cendes}} \emph {et~al.}}]{Cendes2023}%
  \BibitemOpen
  \bibfield  {author} {\bibinfo {author} {\bibfnamefont {Y.}~\bibnamefont
  {{Cendes}}} \emph {et~al.},\ }\bibfield  {title} {\bibinfo {title}
  {{Ubiquitous Late Radio Emission from Tidal Disruption Events}},\ }\href
  {https://doi.org/10.48550/arXiv.2308.13595} {\bibfield  {journal} {\bibinfo
  {journal} {arXiv e-prints}\ ,\ \bibinfo {eid} {arXiv:2309.02275}} (\bibinfo
  {year} {2023})}\BibitemShut {NoStop}%
\bibitem [{\citenamefont {Giannios}\ and\ \citenamefont
  {Metzger}(2011)}]{Giannios2011}%
  \BibitemOpen
  \bibfield  {author} {\bibinfo {author} {\bibfnamefont {D.}~\bibnamefont
  {Giannios}}\ and\ \bibinfo {author} {\bibfnamefont {B.~D.}\ \bibnamefont
  {Metzger}},\ }\bibfield  {title} {\bibinfo {title} {{Radio transients from
  stellar tidal disruption by massive black holes}},\ }\href
  {https://doi.org/10.1111/j.1365-2966.2011.19188.x} {\bibfield  {journal}
  {\bibinfo  {journal} {Mon. Not. Roy. Astron. Soc.}\ }\textbf {\bibinfo
  {volume} {416}},\ \bibinfo {pages} {2102} (\bibinfo {year}
  {2011})}\BibitemShut {NoStop}%
\bibitem [{\citenamefont {{Piran}}\ \emph {et~al.}(2015)\citenamefont {{Piran}}
  \emph {et~al.}}]{Piran2015}%
  \BibitemOpen
  \bibfield  {author} {\bibinfo {author} {\bibfnamefont {T.}~\bibnamefont
  {{Piran}}} \emph {et~al.},\ }\bibfield  {title} {\bibinfo {title}
  {{''Circularization'' vs. Accretion -- What Powers Tidal Disruption
  Events?}},\ }\href {https://doi.org/10.1088/0004-637X/806/2/164} {\bibfield
  {journal} {\bibinfo  {journal} {Astrophys. J.}\ }\textbf {\bibinfo {volume}
  {806}},\ \bibinfo {pages} {164} (\bibinfo {year} {2015})}\BibitemShut
  {NoStop}%
\bibitem [{\citenamefont {van Velzen}\ \emph {et~al.}(2016)\citenamefont {van
  Velzen} \emph {et~al.}}]{van_Velzen2016}%
  \BibitemOpen
  \bibfield  {author} {\bibinfo {author} {\bibfnamefont {S.}~\bibnamefont {van
  Velzen}} \emph {et~al.},\ }\bibfield  {title} {\bibinfo {title} {{A radio jet
  from the optical and x-ray bright stellar tidal disruption flare
  ASASSN-14li}},\ }\href {https://doi.org/10.1126/science.aad1182} {\bibfield
  {journal} {\bibinfo  {journal} {Science}\ }\textbf {\bibinfo {volume}
  {351}},\ \bibinfo {pages} {62} (\bibinfo {year} {2016})}\BibitemShut
  {NoStop}%
\bibitem [{\citenamefont {Matsumoto}\ and\ \citenamefont
  {Piran}(2023)}]{Matsumoto:2022hqp}%
  \BibitemOpen
  \bibfield  {author} {\bibinfo {author} {\bibfnamefont {T.}~\bibnamefont
  {Matsumoto}}\ and\ \bibinfo {author} {\bibfnamefont {T.}~\bibnamefont
  {Piran}},\ }\bibfield  {title} {\bibinfo {title} {{Generalized equipartition
  method from an arbitrary viewing angle}},\ }\href
  {https://doi.org/10.1093/mnras/stad1269} {\bibfield  {journal} {\bibinfo
  {journal} {\mnras}\ }\textbf {\bibinfo {volume} {522}},\ \bibinfo {pages}
  {4565} (\bibinfo {year} {2023})}\BibitemShut {NoStop}%
\bibitem [{\citenamefont {Beniamini}\ \emph {et~al.}(2023)\citenamefont
  {Beniamini}, \citenamefont {Piran},\ and\ \citenamefont
  {Matsumoto}}]{Beniamini2023}%
  \BibitemOpen
  \bibfield  {author} {\bibinfo {author} {\bibfnamefont {P.}~\bibnamefont
  {Beniamini}}, \bibinfo {author} {\bibfnamefont {T.}~\bibnamefont {Piran}},\
  and\ \bibinfo {author} {\bibfnamefont {T.}~\bibnamefont {Matsumoto}},\
  }\bibfield  {title} {\bibinfo {title} {{Swift J1644+57 as an off-axis Jet}},\
  }\href {https://doi.org/10.1093/mnras/stad1950} {\bibfield  {journal}
  {\bibinfo  {journal} {Mon. Not. Roy. Astron. Soc.}\ }\textbf {\bibinfo
  {volume} {524}},\ \bibinfo {pages} {1386} (\bibinfo {year}
  {2023})}\BibitemShut {NoStop}%
\bibitem [{\citenamefont {{Sfaradi}}\ \emph {et~al.}(2024)\citenamefont
  {{Sfaradi}} \emph {et~al.}}]{Sfaradi2023}%
  \BibitemOpen
  \bibfield  {author} {\bibinfo {author} {\bibfnamefont {I.}~\bibnamefont
  {{Sfaradi}}} \emph {et~al.},\ }\bibfield  {title} {\bibinfo {title} {{An
  off-axis relativistic jet seen in the long lasting delayed radio flare of the
  TDE AT 2018hyz}},\ }\href {https://doi.org/10.1093/mnras/stad3717} {\bibfield
   {journal} {\bibinfo  {journal} {\mnras}\ }\textbf {\bibinfo {volume}
  {527}},\ \bibinfo {pages} {7672} (\bibinfo {year} {2024})}\BibitemShut
  {NoStop}%
\bibitem [{\citenamefont {{Alexander}}\ \emph {et~al.}(2016)\citenamefont
  {{Alexander}} \emph {et~al.}}]{Alexander2016}%
  \BibitemOpen
  \bibfield  {author} {\bibinfo {author} {\bibfnamefont {K.~D.}\ \bibnamefont
  {{Alexander}}} \emph {et~al.},\ }\bibfield  {title} {\bibinfo {title}
  {{Discovery of an outflow from radio observations of the tidal disruption
  event ASASSN-14li}},\ }\href {https://doi.org/10.3847/2041-8205/819/2/L25}
  {\bibfield  {journal} {\bibinfo  {journal} {Astrophys. J. Lett.}\ }\textbf
  {\bibinfo {volume} {819}},\ \bibinfo {pages} {L25} (\bibinfo {year}
  {2016})}\BibitemShut {NoStop}%
\bibitem [{\citenamefont {Matsumoto}\ \emph {et~al.}(2022)\citenamefont
  {Matsumoto}, \citenamefont {Piran},\ and\ \citenamefont
  {Krolik}}]{Matsumoto:2021qqo}%
  \BibitemOpen
  \bibfield  {author} {\bibinfo {author} {\bibfnamefont {T.}~\bibnamefont
  {Matsumoto}}, \bibinfo {author} {\bibfnamefont {T.}~\bibnamefont {Piran}},\
  and\ \bibinfo {author} {\bibfnamefont {J.~H.}\ \bibnamefont {Krolik}},\
  }\bibfield  {title} {\bibinfo {title} {{What powers the radio emission in TDE
  AT2019dsg: A long-lived jet or the disruption itself?}},\ }\href
  {https://doi.org/10.1093/mnras/stac382} {\bibfield  {journal} {\bibinfo
  {journal} {Mon. Not. Roy. Astron. Soc.}\ }\textbf {\bibinfo {volume} {511}},\
  \bibinfo {pages} {5085} (\bibinfo {year} {2022})}\BibitemShut {NoStop}%
\bibitem [{\citenamefont {{Cendes}}\ \emph {et~al.}(2021)\citenamefont
  {{Cendes}} \emph {et~al.}}]{Cendes2021}%
  \BibitemOpen
  \bibfield  {author} {\bibinfo {author} {\bibfnamefont {Y.}~\bibnamefont
  {{Cendes}}} \emph {et~al.},\ }\bibfield  {title} {\bibinfo {title} {{Radio
  Observations of an Ordinary Outflow from the Tidal Disruption Event
  AT2019dsg}},\ }\href {https://doi.org/10.3847/1538-4357/ac110a} {\bibfield
  {journal} {\bibinfo  {journal} {\apj}\ }\textbf {\bibinfo {volume} {919}},\
  \bibinfo {eid} {127} (\bibinfo {year} {2021})}\BibitemShut {NoStop}%
\bibitem [{\citenamefont {Hayasaki}\ and\ \citenamefont
  {Yamazaki}(2023)}]{Hayasaki2023}%
  \BibitemOpen
  \bibfield  {author} {\bibinfo {author} {\bibfnamefont {K.}~\bibnamefont
  {Hayasaki}}\ and\ \bibinfo {author} {\bibfnamefont {R.}~\bibnamefont
  {Yamazaki}},\ }\bibfield  {title} {\bibinfo {title} {{Disk
  Wind\textendash{}Driven Expanding Radio-emitting Shell in Tidal Disruption
  Events}},\ }\href {https://doi.org/10.3847/1538-4357/ace35a} {\bibfield
  {journal} {\bibinfo  {journal} {Astrophys. J.}\ }\textbf {\bibinfo {volume}
  {954}},\ \bibinfo {pages} {5} (\bibinfo {year} {2023})}\BibitemShut {NoStop}%
\bibitem [{\citenamefont {Rossi}\ \emph {et~al.}(2002)\citenamefont {Rossi},
  \citenamefont {Lazzati},\ and\ \citenamefont {Rees}}]{Rossi2002}%
  \BibitemOpen
  \bibfield  {author} {\bibinfo {author} {\bibfnamefont {E.}~\bibnamefont
  {Rossi}}, \bibinfo {author} {\bibfnamefont {D.}~\bibnamefont {Lazzati}},\
  and\ \bibinfo {author} {\bibfnamefont {M.~J.}\ \bibnamefont {Rees}},\
  }\bibfield  {title} {\bibinfo {title} {{Afterglow lightcurves, viewing angle
  and the jet structure of gamma-ray bursts}},\ }\href
  {https://doi.org/10.1046/j.1365-8711.2002.05363.x} {\bibfield  {journal}
  {\bibinfo  {journal} {Mon. Not. Roy. Astron. Soc.}\ }\textbf {\bibinfo
  {volume} {332}},\ \bibinfo {pages} {945} (\bibinfo {year}
  {2002})}\BibitemShut {NoStop}%
\bibitem [{\citenamefont {Zhang}\ and\ \citenamefont
  {Meszaros}(2002)}]{Zhang2002}%
  \BibitemOpen
  \bibfield  {author} {\bibinfo {author} {\bibfnamefont {B.}~\bibnamefont
  {Zhang}}\ and\ \bibinfo {author} {\bibfnamefont {P.}~\bibnamefont
  {Meszaros}},\ }\bibfield  {title} {\bibinfo {title} {{Gamma-ray burst
  beaming: a universal configuration with a standard energy reservoir?}},\
  }\href {https://doi.org/10.1086/339981} {\bibfield  {journal} {\bibinfo
  {journal} {Astrophys. J.}\ }\textbf {\bibinfo {volume} {571}},\ \bibinfo
  {pages} {876} (\bibinfo {year} {2002})}\BibitemShut {NoStop}%
\bibitem [{\citenamefont {Alexander}\ \emph {et~al.}(2018)\citenamefont
  {Alexander} \emph {et~al.}}]{Alexander2018}%
  \BibitemOpen
  \bibfield  {author} {\bibinfo {author} {\bibfnamefont {K.~D.}\ \bibnamefont
  {Alexander}} \emph {et~al.},\ }\bibfield  {title} {\bibinfo {title} {{A
  Decline in the X-ray through Radio Emission from GW170817 Continues to
  Support an Off-Axis Structured Jet}},\ }\href
  {https://doi.org/10.3847/2041-8213/aad637} {\bibfield  {journal} {\bibinfo
  {journal} {Astrophys. J. Lett.}\ }\textbf {\bibinfo {volume} {863}},\
  \bibinfo {pages} {L18} (\bibinfo {year} {2018})}\BibitemShut {NoStop}%
\bibitem [{\citenamefont {Ioka}\ and\ \citenamefont
  {Nakamura}(2019)}]{Ioka2019}%
  \BibitemOpen
  \bibfield  {author} {\bibinfo {author} {\bibfnamefont {K.}~\bibnamefont
  {Ioka}}\ and\ \bibinfo {author} {\bibfnamefont {T.}~\bibnamefont
  {Nakamura}},\ }\bibfield  {title} {\bibinfo {title} {{Spectral Puzzle of the
  Off-Axis Gamma-Ray Burst in GW170817}},\ }\href
  {https://doi.org/10.1093/mnras/stz1650} {\bibfield  {journal} {\bibinfo
  {journal} {Mon. Not. Roy. Astron. Soc.}\ }\textbf {\bibinfo {volume} {487}},\
  \bibinfo {pages} {4884} (\bibinfo {year} {2019})}\BibitemShut {NoStop}%
\bibitem [{\citenamefont {{Gottlieb}}\ \emph {et~al.}(2020)\citenamefont
  {{Gottlieb}} \emph {et~al.}}]{Gottlieb2020}%
  \BibitemOpen
  \bibfield  {author} {\bibinfo {author} {\bibfnamefont {O.}~\bibnamefont
  {{Gottlieb}}} \emph {et~al.},\ }\bibfield  {title} {\bibinfo {title} {{The
  structure of weakly magnetized \ensuremath{\gamma}-ray burst jets}},\ }\href
  {https://doi.org/10.1093/mnras/staa2567} {\bibfield  {journal} {\bibinfo
  {journal} {Mon. Not. Roy. Astron. Soc.}\ }\textbf {\bibinfo {volume} {498}},\
  \bibinfo {pages} {3320} (\bibinfo {year} {2020})}\BibitemShut {NoStop}%
\bibitem [{\citenamefont {{Zhang}}\ \emph {et~al.}(2023)\citenamefont {{Zhang}}
  \emph {et~al.}}]{Zhang:2023uei}%
  \BibitemOpen
  \bibfield  {author} {\bibinfo {author} {\bibfnamefont {B.~T.}\ \bibnamefont
  {{Zhang}}} \emph {et~al.},\ }\bibfield  {title} {\bibinfo {title} {{The
  origin of very-high-energy gamma-rays from GRB 221009A: implications for
  reverse shock proton synchrotron emission}},\ }\href
  {https://doi.org/10.48550/arXiv.2311.13671} {\bibfield  {journal} {\bibinfo
  {journal} {arXiv e-prints}\ ,\ \bibinfo {eid} {arXiv:2311.13671}} (\bibinfo
  {year} {2023})}\BibitemShut {NoStop}%
\bibitem [{\citenamefont {{Obayashi}}\ \emph {et~al.}(2024)\citenamefont
  {{Obayashi}} \emph {et~al.}}]{Obayashi2024}%
  \BibitemOpen
  \bibfield  {author} {\bibinfo {author} {\bibfnamefont {K.}~\bibnamefont
  {{Obayashi}}} \emph {et~al.},\ }\bibfield  {title} {\bibinfo {title} {{GRB
  080710: A narrow, structured jet showing a late, achromatic peak in the
  optical and infrared afterglow?}},\ }\href
  {https://doi.org/10.1016/j.jheap.2023.12.001} {\bibfield  {journal} {\bibinfo
   {journal} {JHEAp}\ }\textbf {\bibinfo {volume} {41}},\ \bibinfo {pages} {1}
  (\bibinfo {year} {2024})}\BibitemShut {NoStop}%
\bibitem [{\citenamefont {Berger}\ \emph {et~al.}(2003)\citenamefont {Berger}
  \emph {et~al.}}]{Berger2003}%
  \BibitemOpen
  \bibfield  {author} {\bibinfo {author} {\bibfnamefont {E.}~\bibnamefont
  {Berger}} \emph {et~al.},\ }\bibfield  {title} {\bibinfo {title} {{A common
  origin for cosmic explosions inferred from fireball calorimetry}},\ }\href
  {https://doi.org/10.1038/nature01998} {\bibfield  {journal} {\bibinfo
  {journal} {Nature}\ }\textbf {\bibinfo {volume} {426}},\ \bibinfo {pages}
  {154} (\bibinfo {year} {2003})}\BibitemShut {NoStop}%
\bibitem [{\citenamefont {Racusin}\ \emph {et~al.}(2008)\citenamefont {Racusin}
  \emph {et~al.}}]{Racusin2008}%
  \BibitemOpen
  \bibfield  {author} {\bibinfo {author} {\bibfnamefont {J.~L.}\ \bibnamefont
  {Racusin}} \emph {et~al.},\ }\bibfield  {title} {\bibinfo {title} {{GRB
  080319B: A Naked-Eye Stellar Blast from the Distant Universe}},\ }\href
  {https://doi.org/10.1038/nature07270} {\bibfield  {journal} {\bibinfo
  {journal} {Nature}\ }\textbf {\bibinfo {volume} {455}},\ \bibinfo {pages}
  {183} (\bibinfo {year} {2008})}\BibitemShut {NoStop}%
\bibitem [{\citenamefont {{Sato}}\ \emph {et~al.}(2021)\citenamefont {{Sato}}
  \emph {et~al.}}]{Sato2021}%
  \BibitemOpen
  \bibfield  {author} {\bibinfo {author} {\bibfnamefont {Y.}~\bibnamefont
  {{Sato}}} \emph {et~al.},\ }\bibfield  {title} {\bibinfo {title} {{Off-axis
  jet scenario for early afterglow emission of low-luminosity gamma-ray burst
  GRB 190829A}},\ }\href {https://doi.org/10.1093/mnras/stab1273} {\bibfield
  {journal} {\bibinfo  {journal} {Mon. Not. Roy. Astron. Soc.}\ }\textbf
  {\bibinfo {volume} {504}},\ \bibinfo {pages} {5647} (\bibinfo {year}
  {2021})}\BibitemShut {NoStop}%
\bibitem [{\citenamefont {{Sato}}\ \emph
  {et~al.}(2023{\natexlab{a}})\citenamefont {{Sato}} \emph
  {et~al.}}]{Sato2023a}%
  \BibitemOpen
  \bibfield  {author} {\bibinfo {author} {\bibfnamefont {Y.}~\bibnamefont
  {{Sato}}} \emph {et~al.},\ }\bibfield  {title} {\bibinfo {title}
  {{Synchrotron self-compton emission in the two-component jet model for
  gamma-ray bursts}},\ }\href {https://doi.org/10.1016/j.jheap.2022.12.004}
  {\bibfield  {journal} {\bibinfo  {journal} {JHEAp}\ }\textbf {\bibinfo
  {volume} {37}},\ \bibinfo {pages} {51} (\bibinfo {year}
  {2023}{\natexlab{a}})}\BibitemShut {NoStop}%
\bibitem [{\citenamefont {{Sato}}\ \emph
  {et~al.}(2023{\natexlab{b}})\citenamefont {{Sato}} \emph
  {et~al.}}]{Sato2023b}%
  \BibitemOpen
  \bibfield  {author} {\bibinfo {author} {\bibfnamefont {Y.}~\bibnamefont
  {{Sato}}} \emph {et~al.},\ }\bibfield  {title} {\bibinfo {title}
  {{Two-component jet model for multiwavelength afterglow emission of the
  extremely energetic burst GRB~221009A}},\ }\href
  {https://doi.org/10.1093/mnrasl/slad038} {\bibfield  {journal} {\bibinfo
  {journal} {Mon. Not. Roy. Astron. Soc.}\ }\textbf {\bibinfo {volume} {522}},\
  \bibinfo {pages} {L56} (\bibinfo {year} {2023}{\natexlab{b}})}\BibitemShut
  {NoStop}%
\bibitem [{\citenamefont {Wang}\ \emph {et~al.}(2014)\citenamefont {Wang} \emph
  {et~al.}}]{Wang:2014nwa}%
  \BibitemOpen
  \bibfield  {author} {\bibinfo {author} {\bibfnamefont {J.-Z.}\ \bibnamefont
  {Wang}} \emph {et~al.},\ }\bibfield  {title} {\bibinfo {title}
  {{Quasi-Periodic Variations in X-ray Emission and Long-Term Radio
  Observations: Evidence for a Two-Component Jet in Sw J1644+57}},\ }\href
  {https://doi.org/10.1088/0004-637X/788/1/32} {\bibfield  {journal} {\bibinfo
  {journal} {Astrophys. J.}\ }\textbf {\bibinfo {volume} {788}},\ \bibinfo
  {pages} {32} (\bibinfo {year} {2014})}\BibitemShut {NoStop}%
\bibitem [{\citenamefont {{Mimica}}\ \emph {et~al.}(2015)\citenamefont
  {{Mimica}} \emph {et~al.}}]{Mimica2015}%
  \BibitemOpen
  \bibfield  {author} {\bibinfo {author} {\bibfnamefont {P.}~\bibnamefont
  {{Mimica}}} \emph {et~al.},\ }\bibfield  {title} {\bibinfo {title} {{The
  radio afterglow of Swift J1644+57 reveals a powerful jet with fast core and
  slow sheath}},\ }\href {https://doi.org/10.1093/mnras/stv825} {\bibfield
  {journal} {\bibinfo  {journal} {Mon. Not. Roy. Astron. Soc.}\ }\textbf
  {\bibinfo {volume} {450}},\ \bibinfo {pages} {2824} (\bibinfo {year}
  {2015})}\BibitemShut {NoStop}%
\bibitem [{\citenamefont {{Liu}}\ \emph {et~al.}(2015)\citenamefont {{Liu}},
  \citenamefont {{Pe'er}},\ and\ \citenamefont {{Loeb}}}]{Liu2015}%
  \BibitemOpen
  \bibfield  {author} {\bibinfo {author} {\bibfnamefont {D.}~\bibnamefont
  {{Liu}}}, \bibinfo {author} {\bibfnamefont {A.}~\bibnamefont {{Pe'er}}},\
  and\ \bibinfo {author} {\bibfnamefont {A.}~\bibnamefont {{Loeb}}},\
  }\bibfield  {title} {\bibinfo {title} {{A Two-component Jet Model for the
  Tidal Disruption Event Swift J164449.3+573451}},\ }\href
  {https://doi.org/10.1088/0004-637X/798/1/13} {\bibfield  {journal} {\bibinfo
  {journal} {\apj}\ }\textbf {\bibinfo {volume} {798}},\ \bibinfo {eid} {13}
  (\bibinfo {year} {2015})}\BibitemShut {NoStop}%
\bibitem [{\citenamefont {{Zhou}}\ \emph {et~al.}(2024)\citenamefont {{Zhou}}
  \emph {et~al.}}]{Zhou2023}%
  \BibitemOpen
  \bibfield  {author} {\bibinfo {author} {\bibfnamefont {C.}~\bibnamefont
  {{Zhou}}} \emph {et~al.},\ }\bibfield  {title} {\bibinfo {title} {{AT2022cmc:
  A Tidal Disruption Event with a Two-component Jet in a Bondi-profile
  Circumnuclear Medium}},\ }\href {https://doi.org/10.3847/1538-4357/ad20f3}
  {\bibfield  {journal} {\bibinfo  {journal} {Astrophys. J.}\ }\textbf
  {\bibinfo {volume} {963}},\ \bibinfo {pages} {66} (\bibinfo {year}
  {2024})}\BibitemShut {NoStop}%
\bibitem [{\citenamefont {Teboul}\ and\ \citenamefont
  {Metzger}(2023)}]{Teboul2023}%
  \BibitemOpen
  \bibfield  {author} {\bibinfo {author} {\bibfnamefont {O.}~\bibnamefont
  {Teboul}}\ and\ \bibinfo {author} {\bibfnamefont {B.~D.}\ \bibnamefont
  {Metzger}},\ }\bibfield  {title} {\bibinfo {title} {{A Unified Theory of
  Jetted Tidal Disruption Events: From Promptly Escaping Relativistic to
  Delayed Transrelativistic Jets}},\ }\href
  {https://doi.org/10.3847/2041-8213/ad0037} {\bibfield  {journal} {\bibinfo
  {journal} {Astrophys. J. Lett.}\ }\textbf {\bibinfo {volume} {957}},\
  \bibinfo {pages} {L9} (\bibinfo {year} {2023})}\BibitemShut {NoStop}%
\bibitem [{\citenamefont {{Yuan}}\ \emph
  {et~al.}(2024{\natexlab{a}})\citenamefont {{Yuan}} \emph
  {et~al.}}]{Yuan2024}%
  \BibitemOpen
  \bibfield  {author} {\bibinfo {author} {\bibfnamefont {C.}~\bibnamefont
  {{Yuan}}} \emph {et~al.},\ }\bibfield  {title} {\bibinfo {title} {{Structured
  Jet Model for Multiwavelength Observations of the Jetted Tidal Disruption
  Event AT 2022cmc}},\ }\href {https://doi.org/10.48550/arXiv.2406.11513}
  {\bibfield  {journal} {\bibinfo  {journal} {arXiv e-prints}\ ,\ \bibinfo
  {eid} {arXiv:2406.11513}} (\bibinfo {year} {2024}{\natexlab{a}})}\BibitemShut
  {NoStop}%
\bibitem [{\citenamefont {{Zhang}}\ \emph {et~al.}(2021)\citenamefont {{Zhang}}
  \emph {et~al.}}]{Zhang2021}%
  \BibitemOpen
  \bibfield  {author} {\bibinfo {author} {\bibfnamefont {B.~T.}\ \bibnamefont
  {{Zhang}}} \emph {et~al.},\ }\bibfield  {title} {\bibinfo {title} {{External
  Inverse-Compton Emission from Low-luminosity Gamma-Ray Bursts: Application to
  GRB 190829A}},\ }\href {https://doi.org/10.3847/1538-4357/ac0cfc} {\bibfield
  {journal} {\bibinfo  {journal} {Astrophys. J.}\ }\textbf {\bibinfo {volume}
  {920}},\ \bibinfo {pages} {55} (\bibinfo {year} {2021})}\BibitemShut
  {NoStop}%
\bibitem [{\citenamefont {Murase}\ \emph {et~al.}(2011)\citenamefont {Murase}
  \emph {et~al.}}]{Murase:2010fq}%
  \BibitemOpen
  \bibfield  {author} {\bibinfo {author} {\bibfnamefont {K.}~\bibnamefont
  {Murase}} \emph {et~al.},\ }\bibfield  {title} {\bibinfo {title} {{On the
  Implications of Late Internal Dissipation for Shallow-Decay Afterglow
  Emission and Associated High-Energy Gamma-Ray Signals}},\ }\href
  {https://doi.org/10.1088/0004-637X/732/2/77} {\bibfield  {journal} {\bibinfo
  {journal} {Astrophys. J.}\ }\textbf {\bibinfo {volume} {732}},\ \bibinfo
  {pages} {77} (\bibinfo {year} {2011})}\BibitemShut {NoStop}%
\bibitem [{\citenamefont {{Takahashi}}\ \emph {et~al.}(2022)\citenamefont
  {{Takahashi}} \emph {et~al.}}]{Takahashi2022}%
  \BibitemOpen
  \bibfield  {author} {\bibinfo {author} {\bibfnamefont {K.}~\bibnamefont
  {{Takahashi}}} \emph {et~al.},\ }\bibfield  {title} {\bibinfo {title}
  {{Probing particle acceleration at trans-relativistic shocks with off-axis
  gamma-ray burst afterglows}},\ }\href
  {https://doi.org/10.1093/mnras/stac3022} {\bibfield  {journal} {\bibinfo
  {journal} {\mnras}\ }\textbf {\bibinfo {volume} {517}},\ \bibinfo {pages}
  {5541} (\bibinfo {year} {2022})}\BibitemShut {NoStop}%
\bibitem [{\citenamefont {{Granot}}\ \emph {et~al.}(2002)\citenamefont
  {{Granot}} \emph {et~al.}}]{Granot2002}%
  \BibitemOpen
  \bibfield  {author} {\bibinfo {author} {\bibfnamefont {J.}~\bibnamefont
  {{Granot}}} \emph {et~al.},\ }\bibfield  {title} {\bibinfo {title} {{Off-axis
  afterglow emission from jetted gamma-ray bursts}},\ }\href
  {https://doi.org/10.1086/340991} {\bibfield  {journal} {\bibinfo  {journal}
  {Astrophys. J. Lett.}\ }\textbf {\bibinfo {volume} {570}},\ \bibinfo {pages}
  {L61} (\bibinfo {year} {2002})}\BibitemShut {NoStop}%
\bibitem [{\citenamefont {Matsumoto}\ and\ \citenamefont
  {Metzger}(2023)}]{Matsumoto2023}%
  \BibitemOpen
  \bibfield  {author} {\bibinfo {author} {\bibfnamefont {T.}~\bibnamefont
  {Matsumoto}}\ and\ \bibinfo {author} {\bibfnamefont {B.~D.}\ \bibnamefont
  {Metzger}},\ }\bibfield  {title} {\bibinfo {title} {{Synchrotron afterglow
  model for AT 2022cmc: jetted tidal disruption event or engine-powered
  supernova?}},\ }\href {https://doi.org/10.1093/mnras/stad1182} {\bibfield
  {journal} {\bibinfo  {journal} {Mon. Not. Roy. Astron. Soc.}\ }\textbf
  {\bibinfo {volume} {522}},\ \bibinfo {pages} {4028} (\bibinfo {year}
  {2023})}\BibitemShut {NoStop}%
\bibitem [{\citenamefont {{Anumarlapudi}}\ \emph {et~al.}(2024)\citenamefont
  {{Anumarlapudi}} \emph {et~al.}}]{Anumarlapudi2024}%
  \BibitemOpen
  \bibfield  {author} {\bibinfo {author} {\bibfnamefont {A.}~\bibnamefont
  {{Anumarlapudi}}} \emph {et~al.},\ }\bibfield  {title} {\bibinfo {title}
  {{Radio afterglows from tidal disruption events: An unbiased sample from
  ASKAP RACS}},\ }\href {https://doi.org/10.48550/arXiv.2407.12097} {\bibfield
  {journal} {\bibinfo  {journal} {arXiv e-prints}\ ,\ \bibinfo {eid}
  {arXiv:2407.12097}} (\bibinfo {year} {2024})}\BibitemShut {NoStop}%
\bibitem [{\citenamefont {{Hajela}}\ \emph {et~al.}(2024)\citenamefont
  {{Hajela}} \emph {et~al.}}]{Hajela2024}%
  \BibitemOpen
  \bibfield  {author} {\bibinfo {author} {\bibfnamefont {A.}~\bibnamefont
  {{Hajela}}} \emph {et~al.},\ }\bibfield  {title} {\bibinfo {title} {{Eight
  Years of Light from ASASSN-15oi: Towards Understanding the Late-time
  Evolution of TDEs}},\ }\href {https://doi.org/10.48550/arXiv.2407.19019}
  {\bibfield  {journal} {\bibinfo  {journal} {arXiv e-prints}\ ,\ \bibinfo
  {eid} {arXiv:2407.19019}} (\bibinfo {year} {2024})}\BibitemShut {NoStop}%
\bibitem [{\citenamefont {Blandford}\ and\ \citenamefont
  {Znajek}(1977)}]{Blandford:1977ds}%
  \BibitemOpen
  \bibfield  {author} {\bibinfo {author} {\bibfnamefont {R.~D.}\ \bibnamefont
  {Blandford}}\ and\ \bibinfo {author} {\bibfnamefont {R.~L.}\ \bibnamefont
  {Znajek}},\ }\bibfield  {title} {\bibinfo {title} {{Electromagnetic
  extractions of energy from Kerr black holes}},\ }\href
  {https://doi.org/10.1093/mnras/179.3.433} {\bibfield  {journal} {\bibinfo
  {journal} {Mon. Not. Roy. Astron. Soc.}\ }\textbf {\bibinfo {volume} {179}},\
  \bibinfo {pages} {433} (\bibinfo {year} {1977})}\BibitemShut {NoStop}%
\bibitem [{\citenamefont {McKinney}(2006)}]{McKinney:2006tf}%
  \BibitemOpen
  \bibfield  {author} {\bibinfo {author} {\bibfnamefont {J.~C.}\ \bibnamefont
  {McKinney}},\ }\bibfield  {title} {\bibinfo {title} {{General relativistic
  magnetohydrodynamic simulations of jet formation and large-scale propagation
  from black hole accretion systems}},\ }\href
  {https://doi.org/10.1111/j.1365-2966.2006.10256.x} {\bibfield  {journal}
  {\bibinfo  {journal} {Mon. Not. Roy. Astron. Soc.}\ }\textbf {\bibinfo
  {volume} {368}},\ \bibinfo {pages} {1561} (\bibinfo {year}
  {2006})}\BibitemShut {NoStop}%
\bibitem [{\citenamefont {{Tchekhovskoy}}\ \emph {et~al.}(2011)\citenamefont
  {{Tchekhovskoy}}, \citenamefont {{Narayan}},\ and\ \citenamefont
  {{McKinney}}}]{Tchekhovskoy2011}%
  \BibitemOpen
  \bibfield  {author} {\bibinfo {author} {\bibfnamefont {A.}~\bibnamefont
  {{Tchekhovskoy}}}, \bibinfo {author} {\bibfnamefont {R.}~\bibnamefont
  {{Narayan}}},\ and\ \bibinfo {author} {\bibfnamefont {J.~C.}\ \bibnamefont
  {{McKinney}}},\ }\bibfield  {title} {\bibinfo {title} {{Efficient generation
  of jets from magnetically arrested accretion on a rapidly spinning black
  hole}},\ }\href {https://doi.org/10.1111/j.1745-3933.2011.01147.x} {\bibfield
   {journal} {\bibinfo  {journal} {\mnras}\ }\textbf {\bibinfo {volume}
  {418}},\ \bibinfo {pages} {L79} (\bibinfo {year} {2011})}\BibitemShut
  {NoStop}%
\bibitem [{\citenamefont {Dai}\ \emph {et~al.}(2018)\citenamefont {Dai},
  \citenamefont {McKinney}, \citenamefont {Roth}, \citenamefont
  {Ramirez-Ruiz},\ and\ \citenamefont {Miller}}]{Dai:2018jbr}%
  \BibitemOpen
  \bibfield  {author} {\bibinfo {author} {\bibfnamefont {L.}~\bibnamefont
  {Dai}}, \bibinfo {author} {\bibfnamefont {J.~C.}\ \bibnamefont {McKinney}},
  \bibinfo {author} {\bibfnamefont {N.}~\bibnamefont {Roth}}, \bibinfo {author}
  {\bibfnamefont {E.}~\bibnamefont {Ramirez-Ruiz}},\ and\ \bibinfo {author}
  {\bibfnamefont {M.~C.}\ \bibnamefont {Miller}},\ }\bibfield  {title}
  {\bibinfo {title} {{A unified model for tidal disruption events}},\ }\href
  {https://doi.org/10.3847/2041-8213/aab429} {\bibfield  {journal} {\bibinfo
  {journal} {Astrophys. J. Lett.}\ }\textbf {\bibinfo {volume} {859}},\
  \bibinfo {pages} {L20} (\bibinfo {year} {2018})}\BibitemShut {NoStop}%
\bibitem [{\citenamefont {Dai}\ \emph {et~al.}(2021)\citenamefont {Dai},
  \citenamefont {Lodato},\ and\ \citenamefont {Cheng}}]{Dai_2021}%
  \BibitemOpen
  \bibfield  {author} {\bibinfo {author} {\bibfnamefont {J.~L.}\ \bibnamefont
  {Dai}}, \bibinfo {author} {\bibfnamefont {G.}~\bibnamefont {Lodato}},\ and\
  \bibinfo {author} {\bibfnamefont {R.}~\bibnamefont {Cheng}},\ }\bibfield
  {title} {\bibinfo {title} {The physics of accretion discs, winds and jets in
  tidal disruption events},\ }\bibfield  {journal} {\bibinfo  {journal} {Space
  Science Reviews}\ }\textbf {\bibinfo {volume} {217}},\ \href
  {https://doi.org/10.1007/s11214-020-00747-x} {10.1007/s11214-020-00747-x}
  (\bibinfo {year} {2021})\BibitemShut {NoStop}%
\bibitem [{\citenamefont {Curd}\ and\ \citenamefont
  {Narayan}(2019)}]{Curd:2018qvy}%
  \BibitemOpen
  \bibfield  {author} {\bibinfo {author} {\bibfnamefont {B.}~\bibnamefont
  {Curd}}\ and\ \bibinfo {author} {\bibfnamefont {R.}~\bibnamefont {Narayan}},\
  }\bibfield  {title} {\bibinfo {title} {{GRRMHD simulations of tidal
  disruption event accretion discs around supermassive black holes: jet
  formation, spectra, and detectability}},\ }\href
  {https://doi.org/10.1093/mnras/sty3134} {\bibfield  {journal} {\bibinfo
  {journal} {Mon. Not. Roy. Astron. Soc.}\ }\textbf {\bibinfo {volume} {483}},\
  \bibinfo {pages} {565} (\bibinfo {year} {2019})}\BibitemShut {NoStop}%
\bibitem [{\citenamefont {{Lu}}\ \emph {et~al.}(2023)\citenamefont {{Lu}},
  \citenamefont {{Matsumoto}},\ and\ \citenamefont {{Matzner}}}]{Lu2023}%
  \BibitemOpen
  \bibfield  {author} {\bibinfo {author} {\bibfnamefont {W.}~\bibnamefont
  {{Lu}}}, \bibinfo {author} {\bibfnamefont {T.}~\bibnamefont {{Matsumoto}}},\
  and\ \bibinfo {author} {\bibfnamefont {C.~D.}\ \bibnamefont {{Matzner}}},\
  }\bibfield  {title} {\bibinfo {title} {{Misaligned precessing jets are choked
  by the accretion disk wind}},\ }\bibfield  {journal} {\bibinfo  {journal}
  {arXiv e-prints}\ }\href {https://doi.org/10.48550/arXiv.2310.15336}
  {10.48550/arXiv.2310.15336} (\bibinfo {year} {2023})\BibitemShut {NoStop}%
\bibitem [{\citenamefont {Sironi}\ \emph {et~al.}(2013)\citenamefont {Sironi},
  \citenamefont {Spitkovsky},\ and\ \citenamefont {Arons}}]{Sironi:2013ri}%
  \BibitemOpen
  \bibfield  {author} {\bibinfo {author} {\bibfnamefont {L.}~\bibnamefont
  {Sironi}}, \bibinfo {author} {\bibfnamefont {A.}~\bibnamefont {Spitkovsky}},\
  and\ \bibinfo {author} {\bibfnamefont {J.}~\bibnamefont {Arons}},\ }\bibfield
   {title} {\bibinfo {title} {{The Maximum Energy of Accelerated Particles in
  Relativistic Collisionless Shocks}},\ }\href
  {https://doi.org/10.1088/0004-637X/771/1/54} {\bibfield  {journal} {\bibinfo
  {journal} {Astrophys. J.}\ }\textbf {\bibinfo {volume} {771}},\ \bibinfo
  {pages} {54} (\bibinfo {year} {2013})}\BibitemShut {NoStop}%
\bibitem [{\citenamefont {Lemoine}(2015)}]{Lemoine:2014gca}%
  \BibitemOpen
  \bibfield  {author} {\bibinfo {author} {\bibfnamefont {M.}~\bibnamefont
  {Lemoine}},\ }\bibfield  {title} {\bibinfo {title} {{Non-linear collisionless
  damping of Weibel turbulence in relativistic blast waves}},\ }\href
  {https://doi.org/10.1017/S0022377814000920} {\bibfield  {journal} {\bibinfo
  {journal} {J. Plasma Phys.}\ }\textbf {\bibinfo {volume} {81}},\ \bibinfo
  {pages} {4501} (\bibinfo {year} {2015})}\BibitemShut {NoStop}%
\bibitem [{\citenamefont {Reville}\ and\ \citenamefont
  {Bell}(2014)}]{Reville:2014mta}%
  \BibitemOpen
  \bibfield  {author} {\bibinfo {author} {\bibfnamefont {B.}~\bibnamefont
  {Reville}}\ and\ \bibinfo {author} {\bibfnamefont {A.~R.}\ \bibnamefont
  {Bell}},\ }\bibfield  {title} {\bibinfo {title} {{On the maximum energy of
  shock-accelerated cosmic rays at ultra-relativistic shocks}},\ }\href
  {https://doi.org/10.1093/mnras/stu088} {\bibfield  {journal} {\bibinfo
  {journal} {Mon. Not. Roy. Astron. Soc.}\ }\textbf {\bibinfo {volume} {439}},\
  \bibinfo {pages} {2050} (\bibinfo {year} {2014})}\BibitemShut {NoStop}%
\bibitem [{\citenamefont {Bell}\ \emph {et~al.}(2018)\citenamefont {Bell} \emph
  {et~al.}}]{Bell:2017zzx}%
  \BibitemOpen
  \bibfield  {author} {\bibinfo {author} {\bibfnamefont {A.~R.}\ \bibnamefont
  {Bell}} \emph {et~al.},\ }\bibfield  {title} {\bibinfo {title} {{Cosmic Ray
  Acceleration by Relativistic Shocks: Limits and Estimates}},\ }\href
  {https://doi.org/10.1093/mnras/stx2485} {\bibfield  {journal} {\bibinfo
  {journal} {Mon. Not. Roy. Astron. Soc.}\ }\textbf {\bibinfo {volume} {473}},\
  \bibinfo {pages} {2364} (\bibinfo {year} {2018})}\BibitemShut {NoStop}%
\bibitem [{\citenamefont {Huang}\ \emph {et~al.}(2004)\citenamefont {Huang}
  \emph {et~al.}}]{Huang:2003fwa}%
  \BibitemOpen
  \bibfield  {author} {\bibinfo {author} {\bibfnamefont {Y.~F.}\ \bibnamefont
  {Huang}} \emph {et~al.},\ }\bibfield  {title} {\bibinfo {title}
  {{Rebrightening of XRF 030723: Further evidence for a two - component jet in
  gamma-ray burst}},\ }\href@noop {} {\bibfield  {journal} {\bibinfo  {journal}
  {Astrophys. J.}\ }\textbf {\bibinfo {volume} {605}},\ \bibinfo {pages} {300}
  (\bibinfo {year} {2004})}\BibitemShut {NoStop}%
\bibitem [{\citenamefont {Peng}\ \emph {et~al.}(2005)\citenamefont {Peng},
  \citenamefont {Konigl},\ and\ \citenamefont {Granot}}]{Peng:2004gy}%
  \BibitemOpen
  \bibfield  {author} {\bibinfo {author} {\bibfnamefont {F.}~\bibnamefont
  {Peng}}, \bibinfo {author} {\bibfnamefont {A.}~\bibnamefont {Konigl}},\ and\
  \bibinfo {author} {\bibfnamefont {J.}~\bibnamefont {Granot}},\ }\bibfield
  {title} {\bibinfo {title} {{Two-component jet models of gamma-ray burst
  sources}},\ }\href {https://doi.org/10.1086/430045} {\bibfield  {journal}
  {\bibinfo  {journal} {Astrophys. J.}\ }\textbf {\bibinfo {volume} {626}},\
  \bibinfo {pages} {966} (\bibinfo {year} {2005})}\BibitemShut {NoStop}%
\bibitem [{\citenamefont {Sironi}\ and\ \citenamefont
  {Goodman}(2007)}]{Sironi:2007as}%
  \BibitemOpen
  \bibfield  {author} {\bibinfo {author} {\bibfnamefont {L.}~\bibnamefont
  {Sironi}}\ and\ \bibinfo {author} {\bibfnamefont {J.}~\bibnamefont
  {Goodman}},\ }\bibfield  {title} {\bibinfo {title} {{Production of magnetic
  energy by macroscopic turbulence in GRB afterglows}},\ }\href
  {https://doi.org/10.1086/523636} {\bibfield  {journal} {\bibinfo  {journal}
  {Astrophys. J.}\ }\textbf {\bibinfo {volume} {671}},\ \bibinfo {pages} {1858}
  (\bibinfo {year} {2007})}\BibitemShut {NoStop}%
\bibitem [{\citenamefont {{Inoue}}\ \emph {et~al.}(2011)\citenamefont
  {{Inoue}}, \citenamefont {{Asano}},\ and\ \citenamefont
  {{Ioka}}}]{Inoue2011}%
  \BibitemOpen
  \bibfield  {author} {\bibinfo {author} {\bibfnamefont {T.}~\bibnamefont
  {{Inoue}}}, \bibinfo {author} {\bibfnamefont {K.}~\bibnamefont {{Asano}}},\
  and\ \bibinfo {author} {\bibfnamefont {K.}~\bibnamefont {{Ioka}}},\
  }\bibfield  {title} {\bibinfo {title} {{Three-dimensional Simulations of
  Magnetohydrodynamic Turbulence Behind Relativistic Shock Waves and Their
  Implications for Gamma-Ray Bursts}},\ }\href
  {https://doi.org/10.1088/0004-637X/734/2/77} {\bibfield  {journal} {\bibinfo
  {journal} {\apj}\ }\textbf {\bibinfo {volume} {734}},\ \bibinfo {eid} {77}
  (\bibinfo {year} {2011})}\BibitemShut {NoStop}%
\bibitem [{\citenamefont {{Mizuno}}\ \emph {et~al.}(2011)\citenamefont
  {{Mizuno}} \emph {et~al.}}]{Mizuno2011}%
  \BibitemOpen
  \bibfield  {author} {\bibinfo {author} {\bibfnamefont {Y.}~\bibnamefont
  {{Mizuno}}} \emph {et~al.},\ }\bibfield  {title} {\bibinfo {title}
  {{Magnetic-field Amplification by Turbulence in a Relativistic Shock
  Propagating Through an Inhomogeneous Medium}},\ }\href
  {https://doi.org/10.1088/0004-637X/726/2/62} {\bibfield  {journal} {\bibinfo
  {journal} {\apj}\ }\textbf {\bibinfo {volume} {726}},\ \bibinfo {eid} {62}
  (\bibinfo {year} {2011})}\BibitemShut {NoStop}%
\bibitem [{\citenamefont {{Tomita}}\ \emph {et~al.}(2019)\citenamefont
  {{Tomita}}, \citenamefont {{Ohira}},\ and\ \citenamefont
  {{Yamazaki}}}]{Tomita2019}%
  \BibitemOpen
  \bibfield  {author} {\bibinfo {author} {\bibfnamefont {S.}~\bibnamefont
  {{Tomita}}}, \bibinfo {author} {\bibfnamefont {Y.}~\bibnamefont {{Ohira}}},\
  and\ \bibinfo {author} {\bibfnamefont {R.}~\bibnamefont {{Yamazaki}}},\
  }\bibfield  {title} {\bibinfo {title} {{Weibel-mediated Shocks Propagating
  into Inhomogeneous Electron-Positron Plasmas}},\ }\href
  {https://doi.org/10.3847/1538-4357/ab4a10} {\bibfield  {journal} {\bibinfo
  {journal} {\apj}\ }\textbf {\bibinfo {volume} {886}},\ \bibinfo {eid} {54}
  (\bibinfo {year} {2019})}\BibitemShut {NoStop}%
\bibitem [{\citenamefont {{Lemoine}}\ and\ \citenamefont
  {{Pelletier}}(2010)}]{Lemoine2010}%
  \BibitemOpen
  \bibfield  {author} {\bibinfo {author} {\bibfnamefont {M.}~\bibnamefont
  {{Lemoine}}}\ and\ \bibinfo {author} {\bibfnamefont {G.}~\bibnamefont
  {{Pelletier}}},\ }\bibfield  {title} {\bibinfo {title} {{On electromagnetic
  instabilities at ultra-relativistic shock waves}},\ }\href
  {https://doi.org/10.1111/j.1365-2966.2009.15869.x} {\bibfield  {journal}
  {\bibinfo  {journal} {\mnras}\ }\textbf {\bibinfo {volume} {402}},\ \bibinfo
  {pages} {321} (\bibinfo {year} {2010})}\BibitemShut {NoStop}%
\bibitem [{\citenamefont {Sironi}\ \emph {et~al.}(2015)\citenamefont {Sironi},
  \citenamefont {Keshet},\ and\ \citenamefont {Lemoine}}]{Sironi:2015oza}%
  \BibitemOpen
  \bibfield  {author} {\bibinfo {author} {\bibfnamefont {L.}~\bibnamefont
  {Sironi}}, \bibinfo {author} {\bibfnamefont {U.}~\bibnamefont {Keshet}},\
  and\ \bibinfo {author} {\bibfnamefont {M.}~\bibnamefont {Lemoine}},\
  }\bibfield  {title} {\bibinfo {title} {{Relativistic Shocks: Particle
  Acceleration and Magnetization}},\ }\href
  {https://doi.org/10.1007/s11214-015-0181-8} {\bibfield  {journal} {\bibinfo
  {journal} {Space Sci. Rev.}\ }\textbf {\bibinfo {volume} {191}},\ \bibinfo
  {pages} {519} (\bibinfo {year} {2015})}\BibitemShut {NoStop}%
\bibitem [{\citenamefont {Crumley}\ \emph {et~al.}(2019)\citenamefont
  {Crumley}, \citenamefont {Caprioli}, \citenamefont {Markoff},\ and\
  \citenamefont {Spitkovsky}}]{Crumley:2018kvf}%
  \BibitemOpen
  \bibfield  {author} {\bibinfo {author} {\bibfnamefont {P.}~\bibnamefont
  {Crumley}}, \bibinfo {author} {\bibfnamefont {D.}~\bibnamefont {Caprioli}},
  \bibinfo {author} {\bibfnamefont {S.}~\bibnamefont {Markoff}},\ and\ \bibinfo
  {author} {\bibfnamefont {A.}~\bibnamefont {Spitkovsky}},\ }\bibfield  {title}
  {\bibinfo {title} {{Kinetic simulations of mildly relativistic shocks
  \textendash{} I. Particle acceleration in high Mach number shocks}},\ }\href
  {https://doi.org/10.1093/mnras/stz232} {\bibfield  {journal} {\bibinfo
  {journal} {Mon. Not. Roy. Astron. Soc.}\ }\textbf {\bibinfo {volume} {485}},\
  \bibinfo {pages} {5105} (\bibinfo {year} {2019})}\BibitemShut {NoStop}%
\bibitem [{\citenamefont {Asano}\ \emph {et~al.}(2020)\citenamefont {Asano},
  \citenamefont {Murase},\ and\ \citenamefont {Toma}}]{Asano:2020grw}%
  \BibitemOpen
  \bibfield  {author} {\bibinfo {author} {\bibfnamefont {K.}~\bibnamefont
  {Asano}}, \bibinfo {author} {\bibfnamefont {K.}~\bibnamefont {Murase}},\ and\
  \bibinfo {author} {\bibfnamefont {K.}~\bibnamefont {Toma}},\ }\bibfield
  {title} {\bibinfo {title} {{Probing Particle Acceleration through Broadband
  Early Afterglow Emission of MAGIC Gamma-Ray Burst GRB 190114C}},\ }\href
  {https://doi.org/10.3847/1538-4357/abc82c} {\bibfield  {journal} {\bibinfo
  {journal} {Astrophys. J.}\ }\textbf {\bibinfo {volume} {905}},\ \bibinfo
  {pages} {105} (\bibinfo {year} {2020})},\ \Eprint
  {https://arxiv.org/abs/2007.06307} {arXiv:2007.06307 [astro-ph.HE]}
  \BibitemShut {NoStop}%
\bibitem [{\citenamefont {Baganoff}\ \emph {et~al.}(2003)\citenamefont
  {Baganoff} \emph {et~al.}}]{Baganoff2003}%
  \BibitemOpen
  \bibfield  {author} {\bibinfo {author} {\bibfnamefont {F.~K.}\ \bibnamefont
  {Baganoff}} \emph {et~al.},\ }\bibfield  {title} {\bibinfo {title} {{Chandra
  x-ray spectroscopic imaging of Sgr A* and the central parsec of the
  Galaxy}},\ }\href {https://doi.org/10.1086/375145} {\bibfield  {journal}
  {\bibinfo  {journal} {Astrophys. J.}\ }\textbf {\bibinfo {volume} {591}},\
  \bibinfo {pages} {891} (\bibinfo {year} {2003})}\BibitemShut {NoStop}%
\bibitem [{\citenamefont {Igumenshchev}\ \emph {et~al.}(2000)\citenamefont
  {Igumenshchev}, \citenamefont {Abramowicz},\ and\ \citenamefont
  {Narayan}}]{Igumenshchev2000}%
  \BibitemOpen
  \bibfield  {author} {\bibinfo {author} {\bibfnamefont {I.~V.}\ \bibnamefont
  {Igumenshchev}}, \bibinfo {author} {\bibfnamefont {M.~A.}\ \bibnamefont
  {Abramowicz}},\ and\ \bibinfo {author} {\bibfnamefont {R.}~\bibnamefont
  {Narayan}},\ }\bibfield  {title} {\bibinfo {title} {{Numerical simulations of
  convective accretion flows in three dimensions}},\ }\href
  {https://doi.org/10.1086/312755} {\bibfield  {journal} {\bibinfo  {journal}
  {Astrophys. J. Lett.}\ }\textbf {\bibinfo {volume} {537}},\ \bibinfo {pages}
  {L27} (\bibinfo {year} {2000})}\BibitemShut {NoStop}%
\bibitem [{\citenamefont {Bondi}(1952)}]{Bondi1952}%
  \BibitemOpen
  \bibfield  {author} {\bibinfo {author} {\bibfnamefont {H.}~\bibnamefont
  {Bondi}},\ }\bibfield  {title} {\bibinfo {title} {{On spherically symmetrical
  accretion}},\ }\href {https://doi.org/10.1093/mnras/112.2.195} {\bibfield
  {journal} {\bibinfo  {journal} {Mon. Not. Roy. Astron. Soc.}\ }\textbf
  {\bibinfo {volume} {112}},\ \bibinfo {pages} {195} (\bibinfo {year}
  {1952})}\BibitemShut {NoStop}%
\bibitem [{\citenamefont {{Sun}}\ \emph {et~al.}(2015)\citenamefont {{Sun}},
  \citenamefont {{Zhang}},\ and\ \citenamefont {{Li}}}]{Sun2015}%
  \BibitemOpen
  \bibfield  {author} {\bibinfo {author} {\bibfnamefont {H.}~\bibnamefont
  {{Sun}}}, \bibinfo {author} {\bibfnamefont {B.}~\bibnamefont {{Zhang}}},\
  and\ \bibinfo {author} {\bibfnamefont {Z.}~\bibnamefont {{Li}}},\ }\bibfield
  {title} {\bibinfo {title} {{Extragalactic High-energy Transients: Event Rate
  Densities and Luminosity Functions}},\ }\href
  {https://doi.org/10.1088/0004-637X/812/1/33} {\bibfield  {journal} {\bibinfo
  {journal} {\apj}\ }\textbf {\bibinfo {volume} {812}},\ \bibinfo {eid} {33}
  (\bibinfo {year} {2015})}\BibitemShut {NoStop}%
\bibitem [{\citenamefont {Goodwin}\ \emph {et~al.}(2023)\citenamefont {Goodwin}
  \emph {et~al.}}]{Goodwin:2023qcq}%
  \BibitemOpen
  \bibfield  {author} {\bibinfo {author} {\bibfnamefont {A.~J.}\ \bibnamefont
  {Goodwin}} \emph {et~al.},\ }\bibfield  {title} {\bibinfo {title} {{A
  radio-emitting outflow produced by the tidal disruption event AT2020vwl}},\
  }\href {https://doi.org/10.1093/mnras/stad1258} {\bibfield  {journal}
  {\bibinfo  {journal} {Mon. Not. Roy. Astron. Soc.}\ }\textbf {\bibinfo
  {volume} {522}},\ \bibinfo {pages} {5084} (\bibinfo {year}
  {2023})}\BibitemShut {NoStop}%
\bibitem [{\citenamefont {{Christy}}\ \emph {et~al.}(2024)\citenamefont
  {{Christy}} \emph {et~al.}}]{Christy2024}%
  \BibitemOpen
  \bibfield  {author} {\bibinfo {author} {\bibfnamefont {C.~T.}\ \bibnamefont
  {{Christy}}} \emph {et~al.},\ }\bibfield  {title} {\bibinfo {title} {{The
  Peculiar Radio Evolution of the Tidal Disruption Event ASASSN-19bt}},\ }\href
  {https://doi.org/10.48550/arXiv.2404.12431} {\bibfield  {journal} {\bibinfo
  {journal} {arXiv e-prints}\ ,\ \bibinfo {eid} {arXiv:2404.12431}} (\bibinfo
  {year} {2024})}\BibitemShut {NoStop}%
\bibitem [{\citenamefont {{Mukhopadhyay}}\ \emph {et~al.}(2023)\citenamefont
  {{Mukhopadhyay}}, \citenamefont {{Bhattacharya}},\ and\ \citenamefont
  {{Murase}}}]{MM2023}%
  \BibitemOpen
  \bibfield  {author} {\bibinfo {author} {\bibfnamefont {M.}~\bibnamefont
  {{Mukhopadhyay}}}, \bibinfo {author} {\bibfnamefont {M.}~\bibnamefont
  {{Bhattacharya}}},\ and\ \bibinfo {author} {\bibfnamefont {K.}~\bibnamefont
  {{Murase}}},\ }\bibfield  {title} {\bibinfo {title} {{Multi-messenger
  signatures of delayed choked jets in tidal disruption events}},\ }\href
  {https://doi.org/10.48550/arXiv.2309.02275} {\bibfield  {journal} {\bibinfo
  {journal} {arXiv e-prints}\ ,\ \bibinfo {eid} {arXiv:2309.02275}} (\bibinfo
  {year} {2023})}\BibitemShut {NoStop}%
\bibitem [{VLA()}]{VLA}%
  \BibitemOpen
  \href@noop {} {}\bibinfo {note} {The official website of VLA,
  \url{https://public.nrao.edu/telescopes/vla/}}\BibitemShut {NoStop}%
\bibitem [{Mee()}]{MeerKAT}%
  \BibitemOpen
  \href@noop {} {}\bibinfo {note} {The official website of MeerKAT,
  \url{https://www.sarao.ac.za/science/meerkat/about-meerkat/}}\BibitemShut
  {NoStop}%
\bibitem [{ATC()}]{ATCA}%
  \BibitemOpen
  \href@noop {} {}\bibinfo {note} {The official website of ATCA,
  \url{https://www.narrabri.atnf.csiro.au}}\BibitemShut {NoStop}%
\bibitem [{ngV()}]{ngVLA}%
  \BibitemOpen
  \href@noop {} {}\bibinfo {note} {The official website of ngVLA,
  \url{https://ngvla.nrao.edu/}}\BibitemShut {NoStop}%
\bibitem [{SKA()}]{SKA}%
  \BibitemOpen
  \href@noop {} {}\bibinfo {note} {The official website of SKA,
  \url{https://www.skatelescope.org}}\BibitemShut {NoStop}%
\bibitem [{\citenamefont {Cannizzaro}\ \emph {et~al.}(2021)\citenamefont
  {Cannizzaro} \emph {et~al.}}]{Cannizzaro_2021}%
  \BibitemOpen
  \bibfield  {author} {\bibinfo {author} {\bibfnamefont {G.}~\bibnamefont
  {Cannizzaro}} \emph {et~al.},\ }\bibfield  {title} {\bibinfo {title}
  {{Accretion disc cooling and narrow absorption lines in the tidal disruption
  event AT 2019dsg}},\ }\href {https://doi.org/10.1093/mnras/stab851}
  {\bibfield  {journal} {\bibinfo  {journal} {Mon. Not. Roy. Astron. Soc.}\
  }\textbf {\bibinfo {volume} {504}},\ \bibinfo {pages} {792} (\bibinfo {year}
  {2021})}\BibitemShut {NoStop}%
\bibitem [{XMM()}]{XMM-Newton}%
  \BibitemOpen
  \href@noop {} {}\bibinfo {note} {The official website of {\it XMM-Newton},
  \url{https://sci.esa.int/web/xmm-newton/}}\BibitemShut {NoStop}%
\bibitem [{Cha()}]{Chandra}%
  \BibitemOpen
  \href@noop {} {}\bibinfo {note} {The official website of {\it Chandra},
  \url{https://chandra.harvard.edu}}\BibitemShut {NoStop}%
\bibitem [{Ath()}]{Athena}%
  \BibitemOpen
  \href@noop {} {}\bibinfo {note} {The official website of {\it Athena},
  \url{http://www.the-athena-x-ray-observatory.eu/en}}\BibitemShut {NoStop}%
\bibitem [{eRO()}]{eROSITA}%
  \BibitemOpen
  \href@noop {} {}\bibinfo {note} {The official website of {\it eROSITA},
  \url{https://www.mpe.mpg.de/eROSITA}}\BibitemShut {NoStop}%
\bibitem [{Rub()}]{Rubin}%
  \BibitemOpen
  \href@noop {} {}\bibinfo {note} {The official website of Vera C. Rubin
  Observatory, \url{https://www.lsst.org}}\BibitemShut {NoStop}%
\bibitem [{\citenamefont {Reusch}\ \emph {et~al.}(2022)\citenamefont {Reusch}
  \emph {et~al.}}]{Reusch2022}%
  \BibitemOpen
  \bibfield  {author} {\bibinfo {author} {\bibfnamefont {S.}~\bibnamefont
  {Reusch}} \emph {et~al.},\ }\bibfield  {title} {\bibinfo {title} {{Candidate
  Tidal Disruption Event AT2019fdr Coincident with a High-Energy Neutrino}},\
  }\href {https://doi.org/10.1103/PhysRevLett.128.221101} {\bibfield  {journal}
  {\bibinfo  {journal} {Phys. Rev. Lett.}\ }\textbf {\bibinfo {volume} {128}},\
  \bibinfo {pages} {221101} (\bibinfo {year} {2022})}\BibitemShut {NoStop}%
\bibitem [{\citenamefont {van Velzen}\ \emph {et~al.}(2024)\citenamefont {van
  Velzen} \emph {et~al.}}]{vVelzen2021}%
  \BibitemOpen
  \bibfield  {author} {\bibinfo {author} {\bibfnamefont {S.}~\bibnamefont {van
  Velzen}} \emph {et~al.},\ }\bibfield  {title} {\bibinfo {title}
  {{Establishing accretion flares from supermassive black holes as a source of
  high-energy neutrinos}},\ }\href {https://doi.org/10.1093/mnras/stae610}
  {\bibfield  {journal} {\bibinfo  {journal} {Mon. Not. Roy. Astron. Soc.}\
  }\textbf {\bibinfo {volume} {529}},\ \bibinfo {pages} {2559} (\bibinfo {year}
  {2024})}\BibitemShut {NoStop}%
\bibitem [{\citenamefont {Murase}\ \emph {et~al.}(2020)\citenamefont {Murase}
  \emph {et~al.}}]{Murase:2020lnu}%
  \BibitemOpen
  \bibfield  {author} {\bibinfo {author} {\bibfnamefont {K.}~\bibnamefont
  {Murase}} \emph {et~al.},\ }\bibfield  {title} {\bibinfo {title}
  {{High-Energy Neutrino and Gamma-Ray Emission from Tidal Disruption
  Events}},\ }\href {https://doi.org/10.3847/1538-4357/abb3c0} {\bibfield
  {journal} {\bibinfo  {journal} {Astrophys. J.}\ }\textbf {\bibinfo {volume}
  {902}},\ \bibinfo {pages} {108} (\bibinfo {year} {2020})}\BibitemShut
  {NoStop}%
\bibitem [{\citenamefont {{Hayasaki}}\ and\ \citenamefont
  {{Yamazaki}}(2019)}]{Hayasaki2019}%
  \BibitemOpen
  \bibfield  {author} {\bibinfo {author} {\bibfnamefont {K.}~\bibnamefont
  {{Hayasaki}}}\ and\ \bibinfo {author} {\bibfnamefont {R.}~\bibnamefont
  {{Yamazaki}}},\ }\bibfield  {title} {\bibinfo {title} {{Neutrino Emissions
  from Tidal Disruption Remnants}},\ }\href
  {https://doi.org/10.3847/1538-4357/ab44ca} {\bibfield  {journal} {\bibinfo
  {journal} {\apj}\ }\textbf {\bibinfo {volume} {886}},\ \bibinfo {eid} {114}
  (\bibinfo {year} {2019})}\BibitemShut {NoStop}%
\bibitem [{\citenamefont {Winter}\ and\ \citenamefont
  {Lunardini}(2021)}]{Winter:2020ptf}%
  \BibitemOpen
  \bibfield  {author} {\bibinfo {author} {\bibfnamefont {W.}~\bibnamefont
  {Winter}}\ and\ \bibinfo {author} {\bibfnamefont {C.}~\bibnamefont
  {Lunardini}},\ }\bibfield  {title} {\bibinfo {title} {{A concordance scenario
  for the observed neutrino from a tidal disruption event}},\ }\href
  {https://doi.org/10.1038/s41550-021-01343-x} {\bibfield  {journal} {\bibinfo
  {journal} {Nature Astron.}\ }\textbf {\bibinfo {volume} {5}},\ \bibinfo
  {pages} {472} (\bibinfo {year} {2021})}\BibitemShut {NoStop}%
\bibitem [{\citenamefont {{Yuan}}\ \emph
  {et~al.}(2024{\natexlab{b}})\citenamefont {{Yuan}}, \citenamefont
  {{Winter}},\ and\ \citenamefont {{Lunardini}}}]{Yuan:2024foi}%
  \BibitemOpen
  \bibfield  {author} {\bibinfo {author} {\bibfnamefont {C.}~\bibnamefont
  {{Yuan}}}, \bibinfo {author} {\bibfnamefont {W.}~\bibnamefont {{Winter}}},\
  and\ \bibinfo {author} {\bibfnamefont {C.}~\bibnamefont {{Lunardini}}},\
  }\bibfield  {title} {\bibinfo {title} {{AT2021lwx: Another
  Neutrino-Coincident Tidal Disruption Event with a Strong Dust Echo?}},\
  }\href {https://doi.org/10.48550/arXiv.2401.09320} {\bibfield  {journal}
  {\bibinfo  {journal} {arXiv e-prints}\ ,\ \bibinfo {eid} {arXiv:2401.09320}}
  (\bibinfo {year} {2024}{\natexlab{b}})}\BibitemShut {NoStop}%
\bibitem [{\citenamefont {Wang}\ and\ \citenamefont
  {Liu}(2016)}]{Wang:2015mmh}%
  \BibitemOpen
  \bibfield  {author} {\bibinfo {author} {\bibfnamefont {X.-Y.}\ \bibnamefont
  {Wang}}\ and\ \bibinfo {author} {\bibfnamefont {R.-Y.}\ \bibnamefont {Liu}},\
  }\bibfield  {title} {\bibinfo {title} {{Tidal disruption jets of supermassive
  black holes as hidden sources of cosmic rays: explaining the IceCube TeV-PeV
  neutrinos}},\ }\href {https://doi.org/10.1103/PhysRevD.93.083005} {\bibfield
  {journal} {\bibinfo  {journal} {Phys. Rev. D}\ }\textbf {\bibinfo {volume}
  {93}},\ \bibinfo {pages} {083005} (\bibinfo {year} {2016})}\BibitemShut
  {NoStop}%
\bibitem [{\citenamefont {Senno}\ \emph {et~al.}(2017)\citenamefont {Senno},
  \citenamefont {Murase},\ and\ \citenamefont {Meszaros}}]{Senno:2016bso}%
  \BibitemOpen
  \bibfield  {author} {\bibinfo {author} {\bibfnamefont {N.}~\bibnamefont
  {Senno}}, \bibinfo {author} {\bibfnamefont {K.}~\bibnamefont {Murase}},\ and\
  \bibinfo {author} {\bibfnamefont {P.}~\bibnamefont {Meszaros}},\ }\bibfield
  {title} {\bibinfo {title} {{High-energy Neutrino Flares from X-Ray Bright and
  Dark Tidal Disruption Events}},\ }\href
  {https://doi.org/10.3847/1538-4357/aa6344} {\bibfield  {journal} {\bibinfo
  {journal} {Astrophys. J.}\ }\textbf {\bibinfo {volume} {838}},\ \bibinfo
  {pages} {3} (\bibinfo {year} {2017})}\BibitemShut {NoStop}%
\bibitem [{\citenamefont {Wei}\ \emph {et~al.}(2023)\citenamefont {Wei},
  \citenamefont {Zhang},\ and\ \citenamefont {Murase}}]{Wei:2023rmr}%
  \BibitemOpen
  \bibfield  {author} {\bibinfo {author} {\bibfnamefont {Y.}~\bibnamefont
  {Wei}}, \bibinfo {author} {\bibfnamefont {B.~T.}\ \bibnamefont {Zhang}},\
  and\ \bibinfo {author} {\bibfnamefont {K.}~\bibnamefont {Murase}},\
  }\bibfield  {title} {\bibinfo {title} {{Multiwavelength afterglow emission
  from bursts associated with magnetar flares and fast radio bursts}},\ }\href
  {https://doi.org/10.1093/mnras/stad2122} {\bibfield  {journal} {\bibinfo
  {journal} {Mon. Not. Roy. Astron. Soc.}\ }\textbf {\bibinfo {volume} {524}},\
  \bibinfo {pages} {6004} (\bibinfo {year} {2023})}\BibitemShut {NoStop}%
\bibitem [{\citenamefont {Nava}\ \emph {et~al.}(2013)\citenamefont {Nava} \emph
  {et~al.}}]{Nava:2012hq}%
  \BibitemOpen
  \bibfield  {author} {\bibinfo {author} {\bibfnamefont {L.}~\bibnamefont
  {Nava}} \emph {et~al.},\ }\bibfield  {title} {\bibinfo {title} {Afterglow
  emission in gamma-ray bursts \textendash{} {{I}}. {{Pair-enriched}} ambient
  medium and radiative blast waves},\ }\href
  {https://doi.org/10.1093/mnras/stt872} {\bibfield  {journal} {\bibinfo
  {journal} {Mon. Not. Roy. Astron. Soc.}\ }\textbf {\bibinfo {volume} {433}},\
  \bibinfo {pages} {2107} (\bibinfo {year} {2013})}\BibitemShut {NoStop}%
\bibitem [{\citenamefont {Zhang}(2018)}]{zhangPhysicsGammaRayBursts2018}%
  \BibitemOpen
  \bibfield  {author} {\bibinfo {author} {\bibfnamefont {B.}~\bibnamefont
  {Zhang}},\ }\href {https://doi.org/10.1017/9781139226530} {\emph {\bibinfo
  {title} {The {{Physics}} of {{Gamma-Ray Bursts}}}}},\ \bibinfo {edition}
  {1st}\ ed.\ (\bibinfo  {publisher} {{Cambridge University Press}},\ \bibinfo
  {year} {2018})\BibitemShut {NoStop}%
\bibitem [{\citenamefont {Blandford}\ and\ \citenamefont
  {McKee}(1976)}]{Blandford:1976uq}%
  \BibitemOpen
  \bibfield  {author} {\bibinfo {author} {\bibfnamefont {R.~D.}\ \bibnamefont
  {Blandford}}\ and\ \bibinfo {author} {\bibfnamefont {C.~F.}\ \bibnamefont
  {McKee}},\ }\bibfield  {title} {\bibinfo {title} {{Fluid dynamics of
  relativistic blast waves}},\ }\href {https://doi.org/10.1063/1.861619}
  {\bibfield  {journal} {\bibinfo  {journal} {Phys. Fluids}\ }\textbf {\bibinfo
  {volume} {19}},\ \bibinfo {pages} {1130} (\bibinfo {year}
  {1976})}\BibitemShut {NoStop}%
\bibitem [{\citenamefont {{Sedov}}(1959)}]{Sedov59}%
  \BibitemOpen
  \bibfield  {author} {\bibinfo {author} {\bibfnamefont {L.~I.}\ \bibnamefont
  {{Sedov}}},\ }\href@noop {} {\emph {\bibinfo {title} {{Similarity and
  Dimensional Methods in Mechanics}}}}\ (\bibinfo {year} {1959})\BibitemShut
  {NoStop}%
\bibitem [{\citenamefont {{Taylor}}(1950)}]{Taylor50}%
  \BibitemOpen
  \bibfield  {author} {\bibinfo {author} {\bibfnamefont {G.}~\bibnamefont
  {{Taylor}}},\ }\bibfield  {title} {\bibinfo {title} {{The Formation of a
  Blast Wave by a Very Intense Explosion. II. The Atomic Explosion of 1945}},\
  }\href {https://doi.org/10.1098/rspa.1950.0050} {\bibfield  {journal}
  {\bibinfo  {journal} {Proceedings of the Royal Society of London Series A}\
  }\textbf {\bibinfo {volume} {201}},\ \bibinfo {pages} {175} (\bibinfo {year}
  {1950})}\BibitemShut {NoStop}%
\bibitem [{\citenamefont {Ostriker}\ and\ \citenamefont
  {McKee}(1988)}]{Ostriker:1988zz}%
  \BibitemOpen
  \bibfield  {author} {\bibinfo {author} {\bibfnamefont {J.~P.}\ \bibnamefont
  {Ostriker}}\ and\ \bibinfo {author} {\bibfnamefont {C.~F.}\ \bibnamefont
  {McKee}},\ }\bibfield  {title} {\bibinfo {title} {{Astrophysical
  blastwaves}},\ }\href {https://doi.org/10.1103/RevModPhys.60.1} {\bibfield
  {journal} {\bibinfo  {journal} {Rev. Mod. Phys.}\ }\textbf {\bibinfo {volume}
  {60}},\ \bibinfo {pages} {1} (\bibinfo {year} {1988})}\BibitemShut {NoStop}%
\bibitem [{\citenamefont {van Eerten}\ \emph {et~al.}(2010)\citenamefont {van
  Eerten}, \citenamefont {Zhang},\ and\ \citenamefont
  {MacFadyen}}]{vanEerten:2010zh}%
  \BibitemOpen
  \bibfield  {author} {\bibinfo {author} {\bibfnamefont {H.}~\bibnamefont {van
  Eerten}}, \bibinfo {author} {\bibfnamefont {W.}~\bibnamefont {Zhang}},\ and\
  \bibinfo {author} {\bibfnamefont {A.}~\bibnamefont {MacFadyen}},\ }\bibfield
  {title} {\bibinfo {title} {{Off-Axis Gamma-Ray Burst Afterglow Modeling Based
  On A Two-Dimensional Axisymmetric Hydrodynamics Simulation}},\ }\href
  {https://doi.org/10.1088/0004-637X/722/1/235} {\bibfield  {journal} {\bibinfo
   {journal} {Astrophys. J.}\ }\textbf {\bibinfo {volume} {722}},\ \bibinfo
  {pages} {235} (\bibinfo {year} {2010})}\BibitemShut {NoStop}%
\bibitem [{\citenamefont {Piro}\ and\ \citenamefont
  {Gaensler}(2018)}]{Piro:2018jpl}%
  \BibitemOpen
  \bibfield  {author} {\bibinfo {author} {\bibfnamefont {A.~L.}\ \bibnamefont
  {Piro}}\ and\ \bibinfo {author} {\bibfnamefont {B.~M.}\ \bibnamefont
  {Gaensler}},\ }\bibfield  {title} {\bibinfo {title} {{The Dispersion and
  Rotation Measure of Supernova Remnants and Magnetized Stellar Winds:
  Application to Fast Radio Bursts}},\ }\href
  {https://doi.org/10.3847/1538-4357/aac9bc} {\bibfield  {journal} {\bibinfo
  {journal} {Astrophys. J.}\ }\textbf {\bibinfo {volume} {861}},\ \bibinfo
  {pages} {150} (\bibinfo {year} {2018})}\BibitemShut {NoStop}%
\bibitem [{\citenamefont {Granot}\ and\ \citenamefont
  {Sari}(2002)}]{Granot:2001ge}%
  \BibitemOpen
  \bibfield  {author} {\bibinfo {author} {\bibfnamefont {J.}~\bibnamefont
  {Granot}}\ and\ \bibinfo {author} {\bibfnamefont {R.}~\bibnamefont {Sari}},\
  }\bibfield  {title} {\bibinfo {title} {{The shape of spectral breaks in GRB
  afterglows}},\ }\href {https://doi.org/10.1086/338966} {\bibfield  {journal}
  {\bibinfo  {journal} {Astrophys. J.}\ }\textbf {\bibinfo {volume} {568}},\
  \bibinfo {pages} {820} (\bibinfo {year} {2002})}\BibitemShut {NoStop}%
\end{thebibliography}%

\end{document}